\documentclass[reprint,aps,prb,longbibliography,footinbib,floatfix,nobalancelastpage]{revtex4-2}
\usepackage{amsmath}
\usepackage{amsfonts}
\usepackage{amssymb}
\usepackage{mathtools}
\usepackage{graphicx}
\usepackage[dvipsnames]{xcolor}
\usepackage[colorlinks=True,citecolor=gray,linkcolor=Blue,urlcolor=gray,hypertexnames=false]{hyperref}
\usepackage{hypcap}
\usepackage{yfonts}
\usepackage{bm}
\usepackage{ulem}
\usepackage{dsfont}
\usepackage{braket}
\graphicspath{{./},{./figures/}}
\newcommand\eq[1]{\begin{align}#1\end{align}}

\newcommand{\tabcd}{\theta_{\alpha\beta\gamma\lambda}}
\newcommand\Vabcd[1]{V^{(#1)}_{\alpha\beta\gamma\lambda}}

\newcommand\cbox[1]{\vcenter{\hbox{#1}}}

\newcommand\mytitle{Imprints of information scrambling on eigenstates of a quantum chaotic system}

\definecolor{myBlue}{RGB}{31,119,180}
\definecolor{myOrange}{RGB}{255,127,14}
\definecolor{myGreen}{RGB}{44,160,44}
\definecolor{myRed}{RGB}{214,39,40}
\definecolor{myPurple}{RGB}{148,103,189}

\makeatletter
\def\p@figure{\color{myRed}}
\def\p@equation{\color{myRed}}
\makeatother

\newcommand\new[1]{#1}

\begin{document}

\title{\mytitle}

\author{Bikram Pain}
\email{bikram.pain@icts.res.in}
\affiliation{International Centre for Theoretical Sciences, Tata Institute of Fundamental Research, Bengaluru 560089, India}

\author{Ratul Thakur}
\email{ratul.thakur@icts.res.in}
\affiliation{International Centre for Theoretical Sciences, Tata Institute of Fundamental Research, Bengaluru 560089, India}

\author{Sthitadhi Roy}
\email{sthitadhi.roy@icts.res.in}
\affiliation{International Centre for Theoretical Sciences, Tata Institute of Fundamental Research, Bengaluru 560089, India}

\begin{abstract}
How are the spatial and temporal patterns of information scrambling in locally interacting quantum many-body systems imprinted on the eigenstates of the system's time-evolution operator? We address this question by identifying statistical correlations among sets of minimally four eigenstates that provide a unified framework for various measures of information scrambling. These include operator mutual information and operator entanglement entropy of the time-evolution operator, as well as more conventional diagnostics such as two-point dynamical correlations and out-of-time-ordered correlators. We demonstrate this framework {\new{by obtaining numerical results for Floquet Haar-random circuits and}} by deriving exact results for eigenstate correlations in a minimal model of quantum chaos -- Floquet dual-unitary circuits. These results reveal not only the butterfly effect and the information lightcone, but also finer structures of scrambling within the lightcone. Our work thus shows how the eigenstates of a chaotic system can encode the full spatiotemporal anatomy of quantum chaos, going beyond the descriptions offered by random matrix theory and the eigenstate thermalisation hypothesis.
\end{abstract}

\maketitle

The question of information scrambling in isolated quantum many-body systems is intimately connected to thermalisation of local subsystems.
The eigenstate thermalisation hypothesis (ETH)~\cite{deutsch1991quantum,*deutsch2018eigenstate,srednicki1994chaos,dalessio2016from}, which formalises the latter is a statement about statistical properties of matrix elements of local observables between pairs of eigenstates.
However, more recently, it has been realised that there exist higher-order correlations, beyond the ETH, which are necessary to describe the dynamics of information scrambling~\cite{foini2019eigenstate,chan2019eigenstate,pappalardi2022eigenstate,shi2023local,hahn2023statistical,jindal2024generalized,pain2024entanglement,pappalardi2025fulleigenstate,thakur2025log}. 
A fundamental question thus arises: how is the spatiotemporal structure of information scrambling, in locally interacting systems, encoded in the eigenstates of the time-evolution operator and correlations therein. 

We address this question by deriving analytical relations between eigenstate correlations and entanglement measures of the time-evolution operator between spatially and temporally separated subsystems~\cite{zanardi2001entanglement,prosen2007operator,pizorn2009operator,zhou2017operator,jonay2018coarsegraineddynamicsoperatorstate,mascot2023local}.
The relations we derive, therefore, carry explicitly the imprints of the spatiotemporal structure of information scrambling on the eigenstates of a chaotic, quantum many-body system.
In addition, we also show how various entanglement measures of the time-evolution operator, and hence the eigenstate correlations, are related to more commonly employed diagnostics of information scrambling such as dynamical two-point correlators and out-of-time-ordered correlators (OTOC)~\cite{lieb1972finite,roberts2015diagnosing,maldacena2016bound,hosur2016chaos,aleiner2016microscopic,bohrdt2017scrambling,luitz2017information,kukuljan2017weak,nahum2018operator,keyserlingk2018operator,styliaris2021information}.

Our results therefore provide a unified framework for various measures of information scrambling with the basic ingredient of eigenstate correlations at its heart.
{\new{While fragments of this picture have been studied earlier, such as the connection between OTOCs and and eigenstates correlations~\cite{hahn2023statistical}}, the key contribution of our work is the unified picture which ties together the various measures of information scrambling using the fundamental ingredient of eigenstate
correlations. This also leads to new insights onto hitherto overlooked quantities such as the dynamical
two-point correlations.}

To demonstrate our results concretely, we obtain exact results for the eigenstate correlations for a class of minimal models of maximal quantum chaos, namely Floquet dual-unitary circuits. 
The derivation of the unified framework through eigenstate correlations and exact results for them in class of quantum chaotic systems constitutes the central result of this Letter.  

\begin{figure}
\includegraphics[width=\linewidth]{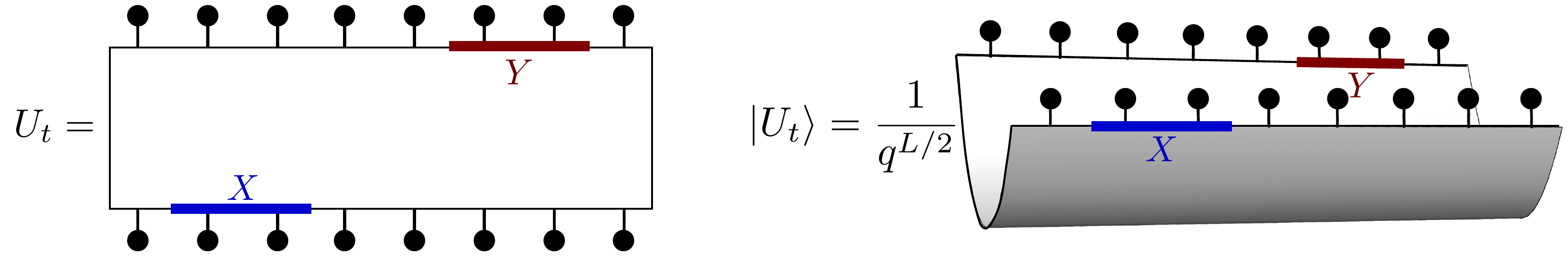}
\caption{The operator $U_t$ acting on $L$ qudits (left) can be equivalently be thought of as a state on $2L$ qudits (right). On the left, the legs at the bottom denote ‘input’ states whereas the those at the top denote the ‘output’ state. Within the state on the doubled system, one can define arbitrary subsystems, exemplified by $X$ and $Y$, which can be separated in both, space and time.}
\label{fig:Ut-op-state}
\end{figure}

The key towards understanding the entanglement properties of the time-evolution operator, $U_t$, is to represent the operator, acting on say $L$ qudits with local Hilbert-space dimension $q$, as a state in the doubled Hilbert-space of $2L$ qudits, $L$ of them each at time ${t=0}$ and at time $t$. 
Formally, the state representation of the operator $U_t = \sum_{i_0,i_t}U_t^{i_0i_t}\ket{i_t}\bra{i_0}$ can be expressed as 
\eq{
	\ket{U_t} = \frac{1}{q^{L/2}}\sum_{i_0,i_t}U_t^{i_0i_t}\ket{i_t \otimes i_0}\,,
	\label{eq:Ut-state}
}
where $\{\ket{i_{0}}\}$ denotes the set of basis states at ${t=0}$ and similarly for $\{\ket{i_{t}}\}$, and the factor of $q^{-L/2}$ ensures normalisation of the doubled state (see Fig.~\ref{fig:Ut-op-state}).

With $U_t$ so mapped to a state one can consider an arbitrary spatiotemporal partition into subsystems of the doubled system. 
Specifically, as shown in Fig.~\ref{fig:Ut-op-state}, consider a subsystem $X$ at time $t=0$, and another subsystem $Y$ at time $t$ such that the two are spatially and temporally separated.
A natural measure for the sptiotemporal strcuture of information then is the second R\'enyi operator mutual information (opMI) between $X$ and $Y$.
\eq{
I_2^{XY}(U_t) = S_2^{X}(U_t)+S_2^{Y}(U_t)-S_2^{X\cup Y}(U_t)\,,
\label{eq:opMI-def}
}
where 
\eq{
	S^{X}_2(U_t) = -\ln {\rm Tr}_X \left[\left({\rm Tr}_{\overline{X}Y\overline{Y}}\ket{U_t}\bra{U_t}\right)^2\right]\,,
	\label{eq:S2-def}
}
is the second R\'enyi operator entanglement entropy (opEE) in $U_t$ between the subsystem $X$ and its complement $\overline{X}Y\overline{Y}$ in the doubled system.
Unitarity of $U_t$ implies that $S_2^{X}(U_t) = |X|\ln q$ and similarly for $Y$, such that $I_2^{XY}(t) = (|X|+|Y|)\ln q -S_2^{X\cup Y}(U_t)$.
This implies that $I_2^{XY}(U_t)$ essentially carries, upto a constant, the entanglement content of $U_t$ between the partitions $X\cup Y$ and $\overline{X}\cup \overline{Y}$.
It is therefore natural to also consider the entanglement between the bipartitions ${X\cup \overline{Y}}$ and ${\overline{X}\cup Y}$ encoded in the opMI $I_2^{X\overline{Y}}(U_t)$ defined similarly to Eq.~\ref{eq:opMI-def}.

We will next relate the opMI to eigenstate correlations. 
In order to do that, let us denote by $U_F$, the generator of time-translation such that $U_t=U_F^t$. 
We will denote the eigenphases and eigenstates of $U_F$ as $\{\theta_\alpha,\ket{\alpha}\}$ such that $U_F = \sum_\alpha e^{-i\theta_\alpha}\ket{\alpha}\bra{\alpha}$.
Note that the second Renyi opEE (Eq.~\ref{eq:S2-def}) necessarily requires four copies of $U_t$, two corresponding to $\ket{U_t}$ and two to $\bra{U_t}$. 
In terms of the eigenstates of $U_F$, this means that the entanglement measures $S_2$ or $I_2$ necessarily involve correlations between {\it quartets} of eigenstates.
Indeed, it has been realised in earlier works~\cite{chan2019eigenstate,shi2023local,hahn2023statistical,jindal2024generalized,pain2024entanglement,thakur2025log} that a description of the dynamics of entanglement necessarily requires correlations between four eigenstates minimally, and thus manifestly goes beyond the paradigm of the ETH.

In order to extract real-space information from the eigenstates, it will be useful to express the eigenstates, given a subsystems $X$, as 
\eq{
	\ket{\alpha} = \sum_{i_X,i_{\overline{X}}}\alpha_{i_Xi_{\overline{X}}}\ket{i_X}\otimes\ket{i_{\overline{X}}}\,,
}
where $\{\ket{i_X}\}$ denotes a set of basis states for subsystem $X$ and similarly for $\{\ket{i_{\overline{X}}}\}$
Using this decomposition, the opMI in Eq.~\ref{eq:opMI-def} can be explicitly written as 
\eq{
	\exp\left[I_2^{XY}(U_t)\right] = q^{|X|+|Y|-2L}F^{XY}(t)\,,
    \label{eq:opMI-FXY}
}
where $F^{XY}$, a sum of dynamical eigenstate correlations each involving four eigenstates, is given by 
\eq{
	F^{XY}(t) = \sum_{\alpha\beta\gamma\lambda} e^{-it\tabcd}\Vabcd{X}\left(\Vabcd{Y}\right)^\ast\,,
	\label{eq:eigcorr-gen-t}
}
with $\Vabcd{X}$ and $\tabcd$ defined, respectively, as

\begin{subequations}
\eq{\Vabcd{X} &= \sum_{i_X,i_{\overline{X}}}\sum_{j_X,j_{\overline{X}}}\alpha_{i_Xi_{\overline{X}}}\beta^\ast_{j_Xi_{\overline{X}}}\gamma^\ast_{i_Xj_{\overline{X}}}\lambda_{j_Xj_{\overline{X}}}\,,
\label{eq:Vabcd}\\
\tabcd & = \theta_\alpha-\theta_\beta-\theta_\gamma+\theta_\lambda\,.
\label{eq:tabcd}
}
\end{subequations}
The relation in Eq.~\ref{eq:opMI-FXY} shows explicitly how the spatiotemporal structure of quantum information, as encoded in the opMI, is directly imprinted onto the eigenstates of a quantum system.
The eigenstate correlation, $F^{XY}(t)$, in Eq.~\ref{eq:eigcorr-gen-t}, can also be equivalently expressed in the frequency domain
\eq{
	\tilde{F}^{XY}(\omega) = \sum_{\alpha\beta\gamma\lambda} \delta(\omega-\tabcd)\Vabcd{X}\left(\Vabcd{Y}\right)^\ast\,,
	\label{eq:eigcorr-gen}
}
which constitutes a correlation between eigenstate amplitudes and the eigenphases of $U_F$ that are dictated by the anatomy of information scrambling.

The opMI defined in Eq.~\ref{eq:opMI-def} can, in fact, be related explicitly to more conventionally used measure of information scrambling. 
The mod-squared two point dynamical correlation between operator $O_X$ in $X$ and $O_Y$ in $Y$ averaged over all operators which form a complete basis in their respective Hilbert spaces is related to the opMI as 
\eq{
\begin{split}
C_{XY}(t)&\equiv\frac{1}{{q^{2(|X|+|Y|)}}}\sum_{O_X,O_Y}\!\!|\braket{O_Y(t)O_X}|^2  \\
&=\frac{\exp[I_2^{XY}(U_t)]}{q^{2(|X|+|Y|)}}=\frac{F^{XY}(t)}{q^{2L+|X|+|Y|}}\,,
\label{eq:CXY}
\end{split}
}
where for the last equality we used Eq.~\ref{eq:opMI-FXY}.
Similarly, the averaged OTOC between operators in $X$ and $Y$ is related to the opMI as
\eq{
\begin{split}
D_{XY}(t)&=\frac{1}{q^{2(|X|+|Y|)}}\sum_{O_X,O_Y}\!\!\braket{(O_Y(t)O_X)^2}\\& = \frac{\exp[I_2^{X\overline{Y}}(U_t)]}{q^{2|X|}}=\frac{F^{X\overline{Y}}(t)}{q^{L+|X|+|Y|}}\,.
\label{eq:DXY}
\end{split}
}
The above relations therefore imply that the eigenstate correlations of the form in Eq.~\ref{eq:eigcorr-gen-t} not only encode the opMI but also conventional measures of information spreading such as two-point correlations and OTOCs.
This concludes the development of a unified framework for understanding the spatiotemporal structure of information scrambling with the eigenstate correlations 
forming the central quantity of study.
{\new{This unified picture constitutes a fundamental backbone of this work which has been graphically summarised in Fig.~\ref{fig:summary}.}



\begin{figure}
\includegraphics[width=0.95\linewidth]{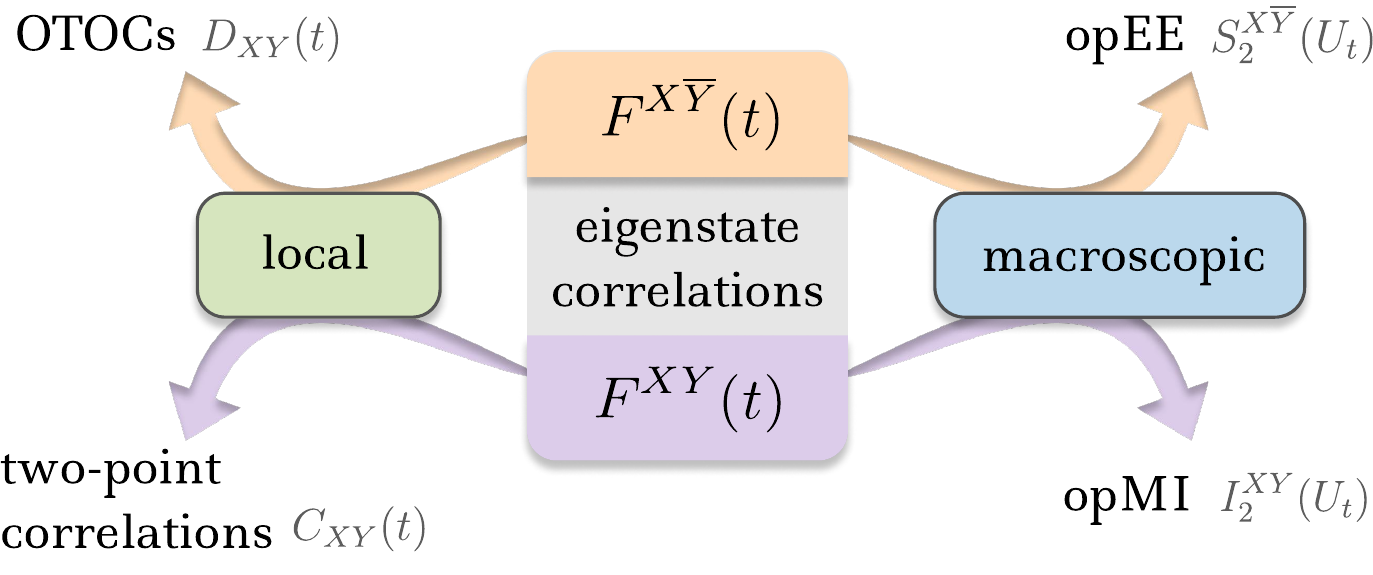}
\caption{Summary of the unified framework for understanding different diagnostics of information scrambling through eigenstate correlations. For $X$ and $Y$ local, the correlations $F^{XY}(t)$ and $F^{X\overline{Y}}(t)$ encode the dynamics of two-point correlations and OTOCs of local operators respectively. For $X$ and $Y$ macroscopic, the correlations govern the dynamics of the opMI and the opEE of appropriate subsystems respectively.}
\label{fig:summary}
\end{figure}

\new{We now present results for these eigenstate correlations in two kinds of Floquet brickwork circuits with $q=2$.
As a generic setting, we first consider Floquet Haar-random circuits (FRUC) for which we present numerical results.
We then specialise to the case of Floquet dual-unitary circuits (FDUC)  for which we obtain exact results. 
For Floquet brickwork circuits, the Floquet unitary is given by 
$U_F = \left(\otimes_{i=1}^{L/2}w_{2i-1,2i})\right)\left(\otimes_{i=1}^{L/2}w_{2i,2i+1}\right)\,.
$
For FRUCs, the $w$ are two-qubit Haar-random gates wheres as for FDUCs}}
the $w$-gates are parametrised as 
\eq{
w = e^{i\phi}(u_+\otimes u_-)\cdot {\cal U}[J] \cdot (v_+\otimes v_-)\,,
\label{eq:w-gate-def}
}
with $\phi,J\in \mathbb{R}$, $u_\pm,v_\pm\in {\rm SU}(2)$, and 
\eq{
\!\!\!{\cal U}[J]\!=\!\exp\!\left[-i\frac{\pi}{4}\left(\sigma^x\!\otimes\!\sigma^x+\sigma^y\!\otimes\!\sigma^y\right)-iJ\sigma^z\!\otimes\!\sigma^z\right]\,.
\label{eq:w-gate}
}
Such a parametrisation leads to dual-unitarity which means that the circuit remains unitary if viewed `sideways' by exchanging space and time. 
These properties have facilitated an array of exact results for DU circuits despite them being non-integrable and chaotic~\cite{bertini2018exactSFF,bertini2019entanglementspreading,bertini2019exactcorrelations,bertini2020opEE,Piroli2020exactdynamics,claeys2020maximumvelocity,fritzsch2021ETHDU,claeys2021ergodicandnonergodic,aravinda2021DUtobernouli,claeys2024operator,Fritzsch2025eigenstate,Foligno2025entanglementof,bertini2025exactlysolvablemanybodydynamics}. 
However, unlike much of previous work which considers random DU circuits, we consider {Floquet} DU circuits which allows us to discuss correlations between eigenstates and eigenvalues of the Floquet unitary $U_F$.
\new{However, it must be noted that FDUCs present significant technical challenges over random DU circuits due to the correlations between gates at different times.}

We consider two specific settings, (i) subsystems $X$ and $Y$ are single sites separated by a spatial distance $r$, and (ii) $X$ and $Y$ are macroscopic subsystems again separated by distance $r$.
For FDUCs, we will consider both $X$ and $Y$ to be located on even sites in the former and the boundaries of $X$ and $Y$ to also be located on even sites in the latter (results for odd-odd and odd-even sites follow similarly with some subtle differences~\cite{supp}. Also, without loss of generality, we will consider $Y$ to be located entirely to the right of $X$.

\new{Results for the eigenstate correlations for both FRUCs and FDUCs are shown in Fig.~\ref{fig:numerical}. We focus on $\tilde{F}^{XY}$ with $X,Y$ local and $\tilde{F}^{X\overline{Y}}$ with $X,Y$ macroscopic corresponding to two-point correlations and opEE respectively~\footnote{The case of $\tilde{F}^{X\overline{Y}}$ with $X,Y$ local corresponds to the OTOC which was studied in Ref.~\cite{hahn2023statistical} and is expected to be similar in the two kinds of circuits.}.
Physically, the results can be understood as follows. 
Two-point functions vanish identically outside the lightcone.
Inside, for FRUCs they decay exponentially in both space and time whereas for FDUCs, they survive solely on the lightcone decaying exponentially with $t$.
Consequently,  $\tilde{F}^{XY}$ exhibits oscillatory behaviour in $\omega$: its amplitude decays exponentially with $r$ and its frequency grows linearly with $r$, accompanied by a weak (no) damping for FRUCs (FDUCs). 
Similarly, the opEE vanishes outside the lightcone and grows linearly inside, yielding damped oscillations in  
$\tilde{F}^{X\overline{Y}}$ with a with a frequency $\propto r$. 
The most important takeaway from the results is the qualitative similarity between those for FRUCs and FDUCs.
This motivates us to obtain exact results for the eigenstate correlations for FDUCs which constitutes much of what follows.

\begin{figure}
\includegraphics[width=\linewidth]{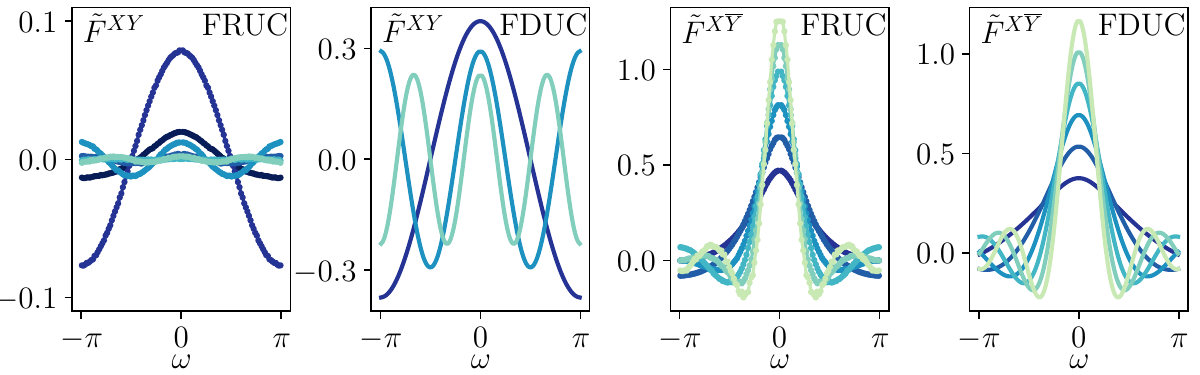}
 \caption{Eigenstate correlation, $\tilde{F}^{XY}(\omega)$ for $X,Y$ local and $\tilde{F}^{X\overline{Y}}(\omega)$ for $X,Y$ macroscopic for both the FRUCs and FDUCs. Results for the former were obtained using exact diagonalisation for $L=12$ whereas those for the latter were obtained analytically in the thermodynamic limit (see Eq.~\ref{eq:FXYt-local-res} and Eq.~\ref{eq:FXYbar-nonlocal-res}) with $J=1$. Different colours denote different $r=1,2,\cdots,6$ with brighter colours denoting larger $r$.} 
\label{fig:numerical}
\end{figure}

}

\begin{figure}[!b]
    \includegraphics[width=\linewidth]{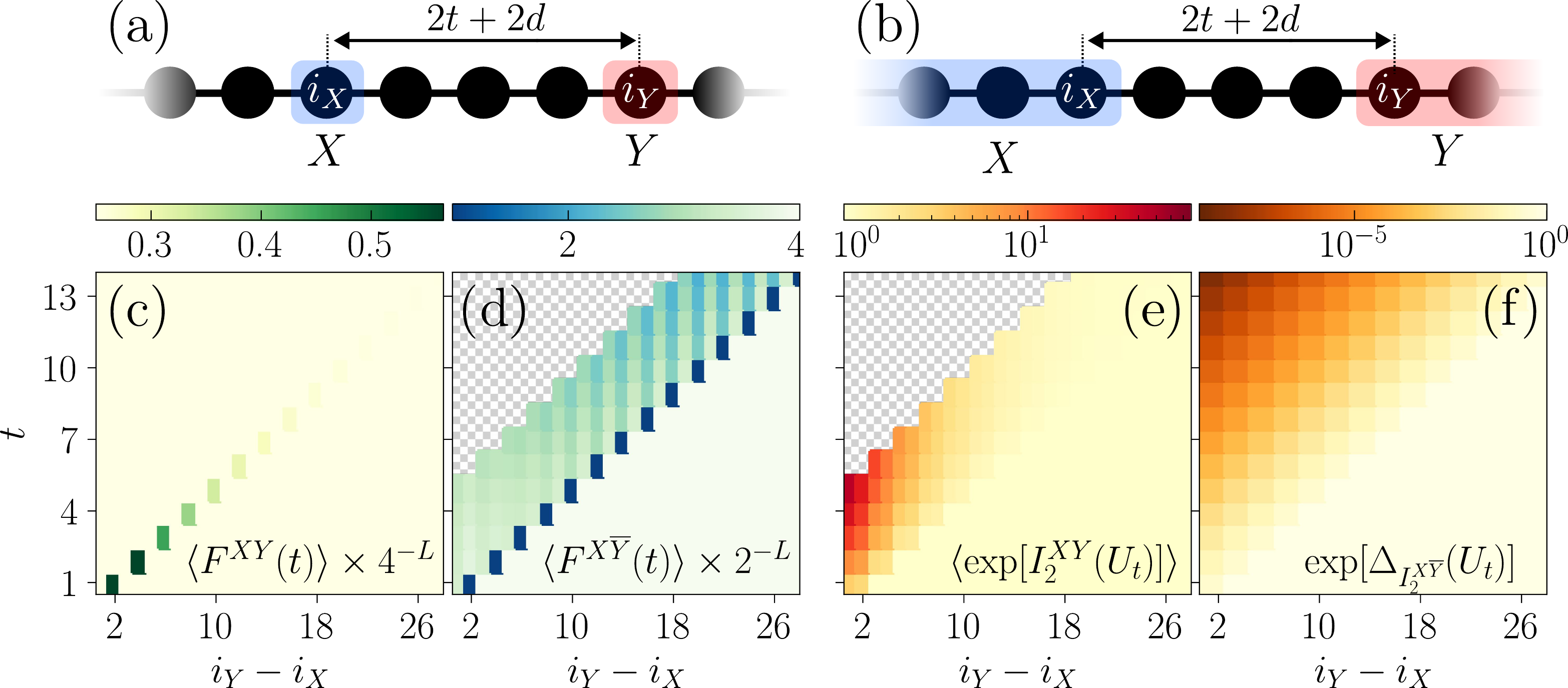}
    \caption{Summary of the eigenstate correlations as heatmaps in the space-time plane. Schematics showing the settings where the subsystems $X$ and $Y$ are (a) local and (b) macroscopic. Panels (c) and (d) show the appropriately rescaled $F^{XY}$ and $F^{X\overline{Y}}$ for $X$ and $Y$ local, which in turn corresponds to the two-point correlations and OTOCs. Panels (e) and (f) correspond to $X$, $Y$ macroscopic where it is more revealing to present the results in terms of the opMIs, $I_2^{XY}$ and $I_2^{X\overline{Y}}$. The hatched regions in (d), (e) correspond to values of $d$ where the transfer matrix is too large for the analysis to be feasible.}
    \label{fig:lightcones}
\end{figure}

These results are obtained through a combination of circuit-diagrammatic techniques and analysis of emergent transfer matrices~\cite{supp}.
The brickwork geometry of the circuit imposes a natural lightcone velocity of $v=2$. 
We will therefore find it convenient to parametrise the distance between $X$ and $Y$ relative to the lightcone ray. 
As such, when $X$ and $Y$ are single sites indexed by $i_X$ and $i_Y$, we will consider $i_Y = i_X + 2t+2d$.
On the other hand, for $X$ and $Y$ macroscopic, we will consider the rightmost site of $X$ and the leftmost site of $Y$ to have indices which differ by $2t+2d$. These two settings are shown graphically in Fig.~\ref{fig:lightcones}(a)-(b). 
With these notations in place, we now present and discuss the results. 

In the case where $X$ and $Y$ are single sites, $F^{XY}(t)$ averaged over the single-site Haar random matrices $(u_\pm,v_\pm)$ parametrising $U_F$ evaluates to 
\eq{
	\braket{F^{XY}(t)} &= 4^{L-1}\left[1+3\Lambda^{2t}\delta_{d0}\right]\,,
	\label{eq:FXYt-local-res}
}
where $\Lambda = [2-\cos(4J)]/3$.
The form of this result stems from the fact that for $d=0$, $\braket{F^{XY}(t)}$ can effectively written in terms of a one-dimensional tensor network, built out of repeated applications of a transfer matrix (see Eq.~\ref{eq:FXYt-local-circs}) which has two non-zero eigenvalues, 4 and $4\Lambda$~\cite{supp}.
The result implies that for $Y$ inside and outside the lightcone emanating from $X$, $F^{XY}$ is a constant for all times whereas on the lightcone, the eigenstate correlation decays with time exponentially with a rate $2|\ln\Lambda|$.
Equivalently, the relation in Eq.~\ref{eq:opMI-FXY} implies that that inside and outside the lightcone the opMI is identically zero whereas on the lightcone the opMI decays exponentially in time at late times as $\ln(1+3\Lambda^{2t})\sim \Lambda^{2t}$. This constitutes our first demonstration of how the eigenstate correlations encode the spatiotemporal dynamics of information in these systems.

Using this result for the opMI in Eq.~\ref{eq:CXY}, we also have the result that $C_{XY}(t)=1/16$ inside and outside the lightcone whereas on the lightcone, it decays exponentially with time and saturates to a value of 1/16 at $t\to\infty$.
This reflects the known result that in dual-unitary circuits, all the non-trivial two-point dynamical correlations (except the trivial one corresponding to $O_X=\mathbb{I}$ and $O_Y=\mathbb{I}$) are identically zero away from the lightcone whereas they decay exponentially on it~\cite{bertini2019exactcorrelations,claeys2020maximumvelocity}.

From the point of view of correlations between eigenstates and eigenvalues, {\it \`a la} Eq.~\ref{eq:eigcorr-gen}, the above result implies that for $X$ and $Y$ at a fixed distance $r$ from each other, we have 
\eq{
\braket{\tilde{F}^{XY}_{\rm local}(r,\omega)} = 4^{L-1}\left[\delta(\omega) + \frac{3\Lambda^r}{2\pi} \cos\left(\frac{\omega r}{2}\right)\right]\,,
}
which explicitly shows how the local structure of the dynamics is encoded in the eigenstate correlations.

Turning to $F^{X\overline{Y}}(t)$ in the same setting, a similar analysis leads to the result, 
\eq{
	\braket{F^{X\overline{Y}}(t)} &= 2^L\times \begin{cases}
    \frac{7}{4}(1+c_d\Gamma_d^{2t}+\cdots)\,; & d<0\\
    1\,; & d=0\\
    4\,; & d>0
    \end{cases}\,,
	\label{eq:FXYbart-local-res}
}
where ${0<\Gamma_d<1}$ decreases with increasing $|d|$ suggesting the correlation decays faster the further inside $Y$ is of the lightcone, and the ellipsis denote further subleading corrections.
Note that our convention mandates that the first line of the above equation is valid only for $t\geq|d|$~\footnote{This is simply due to the fact that for $i_Y$ to lie at distance $2|d|$ inside the lightcone, the latter must spread to distance $2|d|$ at least which requires $t>|d|$.}.
In this case, the tensor network for $d<0$ is no longer one-dimensional but has the geometry of a ribbon of width $|d|+1$~\cite{supp}. The corresponding transfer matrix when written as a rank-2 tensor, is a $4^{2|d|+1}$-dimensional matrix with a leading eigenvalue of 2 and the first subleading eigenvalue $2\Gamma_d$ which we evaluate numerically for a few values of $d$~\cite{supp}.

The relation in Eq.~\ref{eq:DXY} implies that $D_{XY}=1$ outside the lightcone which follows trivially from the fact that $O_X(t)$ and $O_Y$ commute in that case.
On the other hand, inside the lightcone $D_{XY}$ decays exponentially in time and saturates to 7/16; the saturation can be understood simply from the fact that if either or both of $O_X$ and $O_Y$ are $\mathbb{I}$ (7 combinations out of the 16 possible), the OTOC is trivially 1 at all times.
More interestingly, the result indicates that the rate of the exponential decay of the non-trivial OTOCs depends on the distance from the edge of the lightcone and not on the velocity of the ray.
This provides a finer characterisation of the structure of OTOCs beyond velocity-dependent Lyapunov exponents~\cite{khemani2018velocity} which for systems with finite local Hilbert-space dimension is not well defined inside the lightcone due to the absence of extended exponential regime in time.
The result in Eq.~\ref{eq:FXYbart-local-res} therefore constitutes a second instance where eigenstate correlations demonstrably encode the spatiotemporal structure of information scrambling.

We next consider the second setting mentioned above, where the subsystem $X$ and $Y$ are macroscopic subsystems. 
Since, in this case, the eigenstate correlations are related to the two-point dynamical correlations and OTOCs of operators which themselves have macroscopic support, we find it more natural to express the results in terms of the opMI.
The relation in Eq.~\ref{eq:opMI-FXY} and the fact that $|X|+|Y| = L-2(t+d)+1$, implies that it is appropriate to study the rescaled eigenstate correlation $\braket{F^{XY}(t)}/2^{L+2(t+d)-1}$ which is nothing but $\braket{e^{I_2^{XY}(U_t)}}$. 
The circuit diagrams and the resulting transfer matrix~\cite{supp} show that this evaluates to
\eq{
\braket{e^{I_2^{XY}(U_t)}}=\begin{cases}
    1+\frac{3(|d|+1)}{\Lambda^{\vert d \vert}}\Lambda^{2t}+\cdots\,; & \!\!\!\! d\leq  0\\
    1\,; & \!\!\!\!d>0
    \end{cases}\,,
	\label{eq:FXYt-nonlocal-res}
}
where, as before, $\Lambda = [2-\cos(4J)]/3$ and the ellipsis denotes further subleading corrections.
The result implies that the opMI between $X$ and $Y$ is identically zero if they fall entirely outside the lightcone whereas inside the lightcone while the opMI is non-zero, it decays exponentially with $t$ at long times with a rate which is independent of $d$.
Another fallout of the above result is that along a given velocity ray, defined by $i_Y=i_X+vt$, the opMI is given by $I_2^{XY}(U_t)\sim e^{-\gamma(v)t}$ at long times where $\gamma(v)=|\ln\Lambda|(1+v/2)$ for $v\leq 2$ and $\gamma(v)\to\infty$ for $v>2$ indicating an explicit velocity dependence in the eigenstate correlation. 

In considering $F^{X\overline{Y}}$ or $I_2^{X\overline{Y}}$, note that the subsystems $X$ and $\overline{Y}$ have finite spatial overlap. 
This leads to a finite $I_2^{X\overline{Y}}$ even if $d>0$.
Equivalently, already at time $t=0$, $I_2^{X\overline{Y}}(U_0)=2|X|\ln 2$ with $U_0=\mathbb{I}$.
This suggests that the natural quantity to study is $\Delta_{I_2^{X\overline{Y}}(U_t)}\equiv I_2^{X\overline{Y}}(U_t)-I_2^{X\overline{Y}}(U_0)$.
As before, using Eq.~\ref{eq:opMI-FXY} it can be shown that $\exp[\Delta_{I_2^{X\overline{Y}}(U_t)}] = F^{X\overline{Y}}(t)/2^{2(L-t-d)+1}$.
Evaluating $F^{X\overline{Y}}(t)$ using the circuits diagrams~\cite{supp}, we obtain
\eq{
	\exp\left[\Delta_{I_2^{X\overline{Y}}(U_t)}\right] &=  \begin{cases}
    2^{2d-2}\,; & d\leq 0\\
    1\,; & d>0
    \end{cases}\,.
	\label{eq:FXYbart-nonlocal-res}
}
The result above shows that $I_2^{X\overline{Y}}(U_t)$ sticks to its $t=0$ value if $X$ and $Y$ fall entirely outside each others lightcone whereas if they are within the lightcone the opMI jumps by an amount, $-2(|d|+1)\ln 2$.
It is more revealing to note that $-\Delta_{I_2^{X\overline{Y}}(U_t)}$ is the opEE of $X\cup \overline{Y}$ relative to its $t=0$ value, 
\eq{
-\Delta_{I_2^{X\overline{Y}}(U_t)}=S_2^{X\cup \overline{Y}}(U_t) - S_2^{X\cup \overline{Y}}(U_0)\,.\nonumber
}
Considering $i_Y=i_X+vt$ to lie on a velocity ray from $i_X$, the result implies that $S_2^{X\overline{Y}}(U_t) = S_2^{X\overline{Y}}(U_0)$ for $v>2$ whereas for $v\leq 2$, 
\[S_2^{X\overline{Y}}(U_t) = S_2^{X\overline{Y}}(U_0) + 2\ln2+ t(2-v)\ln 2\,,\]
which makes the linear in time growth of the opEE for velocity rays inside the lightcone explicit.
The results in Eq.~\ref{eq:FXYt-nonlocal-res} and Eq.~\ref{eq:FXYbart-nonlocal-res} show two more instances of how the spatiotemporal structure of information scrambling is manifestly imprinted on eigenstate correlations, also in the case where the subsystems are macroscopically large. 

Using the relation between opMI and eigenstate correlations, \eqref{eq:opMI-FXY}, the above result implies for the latter with $X$ and $Y$ macroscopic, separated by a fixed $r$, that,
\eq{
\tilde{F}^{X\overline{Y}}_{\rm macro}(r,\omega)\!=\!\frac{2^{2L}}{2^r\pi}\left[\frac{\sin(\frac{\omega r}{2})}{\omega}\!+\! \frac{\cos(\frac{\omega r}{2})\ln 4-\omega\sin(\frac{\omega r}{2})}{4(\omega^2+4\ln^2 2)}\right]\,,\label{eq:FXYbar-nonlocal-res}
}
which again presents a concrete manifestation of the spatiotemporal structure of information scrambling on the eigenstate correlation in Eq.~\ref{eq:eigcorr-gen}.

While we provided explicit expressions for the eigenstate correlations in various settings, we also present them in Fig.~\ref{fig:lightcones} as appropriately normalised heatmaps in the space-time plane, offering a clear visual summary of the results. These graphical representations clearly reveal the signatures of the scrambling lightcone and the structure of eigenstate correlations within it.

We close with a brief summary and concluding remarks. This work has two main results. First, we establish a unified framework for understanding the spatiotemporal structure of information scrambling in chaotic quantum systems through the lens of dynamical eigenstate correlations of the time-evolution operator. In particular, we show how such correlations encode the anatomy of information scrambling, as quantified by the opMI and opEE of the time-evolution operator -- quantities that are themselves closely related to the dynamics of two-point functions and OTOCs. This framework, summarized in Fig.~\ref{fig:summary}, provides a concrete answer to the question posed at the outset: what are the imprints of scrambling on the eigenstates of the time-evolution operator? The answer lies in four-eigenstate correlations that go beyond the predictions of both the ETH and random matrix theory.
As an aside, an interesting upshot is that two-point dynamical correlations of operators with both local and macroscopic support can be used to reconstruct local OTOCs.

The second main result of this work is a set of exact expressions for eigenstate correlations in Floquet dual-unitary circuits {\new{and their similarities to Floquet Haar-random circuits}}. These results make manifest the structure of the scrambling lightcone as well as finer features within it. For local subsystems, the correlations reveal that, inside the lightcone, OTOCs decay exponentially in time, with a rate determined by the distance from the lightcone. For macroscopic subsystems, the correlations exhibit a velocity dependence of the opMI and opEE of the time-evolution operator. Moreover, these results provide insight into the autocorrelations of nonlocal (multisite) operators, which are expected to encode information about the (partial) spectral form factor in such systems~\cite{bertini2018exactSFF,Fritzsch2025eigenstate,yoshimura2025operator}. We leave a detailed exploration of this connection to future work.

While we focussed on chaotic Floquet dual-unitary circuits for exact results, the framework developed in this work (and summarised in Fig.~\ref{fig:summary}) is completely general. 
It will therefore be interesting to study the fate of these correlations for dual-unitary circuits with varying levels of ergodicity~\cite{claeys2021ergodicandnonergodic,alves2025probes}.
Of particular interest are chaotic systems with conservation laws where the eigenstate correlations may provide a microscopic explanation for anomalous decay of correlations~\cite{mcculloch2025subexponential} as well as the anomalous growth of higher R\'enyi etropies of entanglement~\cite{rakovszky2019subballistic,zhou2020diffusive,znidarivc2020entanglement,huang2020dynamics}.

\begin{acknowledgments}
We used the {\texttt{ITensor}} library~\cite{itensor} to construct and analyse the averaged transfer matrices.
We thank S. Mandal for useful discussions. This work was supported by the Department of Atomic Energy, Government of India un- der Project No. RTI4001, by SERB-DST, Government of India under Grant No. SRG/2023/000858 and by a Max Planck Partner Group grant between ICTS-TIFR, Bengaluru and MPIPKS, Dresden.
\end{acknowledgments}

\bibliography{refs}

\clearpage

\clearpage

\newpage

\setcounter{equation}{0}
\setcounter{figure}{0}
\setcounter{page}{1}
\renewcommand{\theequation}{S\arabic{equation}}
\renewcommand{\thefigure}{S\arabic{figure}}
\renewcommand{\thesection}{SM \Roman{section}}
\renewcommand{\thepage}{S\arabic{page}}

\onecolumngrid

\begin{center}
    {\bf Supplementary Material: \mytitle}\\
    Bikram Pain, Ratul Thakur, and Sthitadhi Roy\\
    {\small{\it International Centre for Theoretical Sciences, Tata Institute of Fundamental Research, Bengaluru 560089, India}}
\end{center}

\section{Diagrammatics for eigenstate correlations}

In the main text, we mentioned that the exact results for the eigenstate correlations can be obtained using a combination of circuit-diagrammatic calculations and analysis of emergent transfer matrices.
Here we present the diagrammatic notations and rules which are useful for obtaining the results. 
As is clear from the form of $F^{XY}(t)$, we need four copies of the unitary time-evolution operator, two corresponding to $U_t$ and two corresponding to $U_t^\ast$.
We therefore need a notation for the four copies, as well as the two kinds of contractions at a leg carrying 4 indices with each of the index taking $q=2$ values. 
Denoting a unitary operator $w$ acting on two sites as a white box, (and $w^\ast$ with a light grey box) we define the notation for the four-copy picture  and the two kinds of contractions as
\eq{
	\cbox{\includegraphics[width=0.05\linewidth]{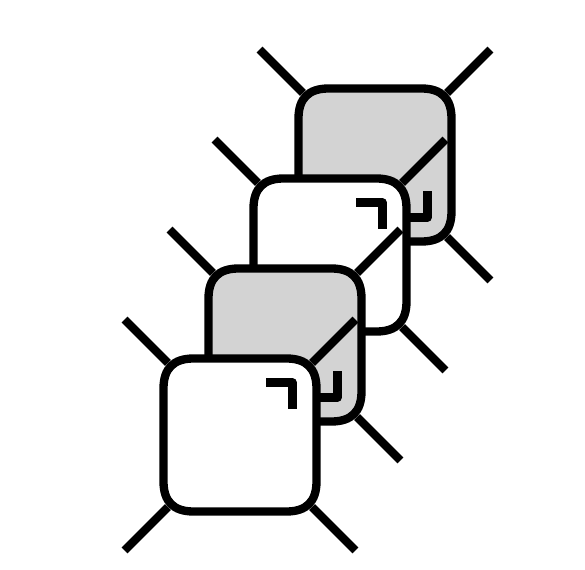}} := \cbox{\includegraphics[width=0.04\linewidth]{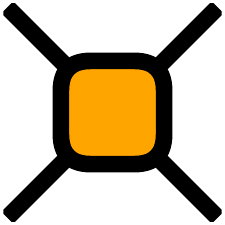}}\,,\quad
    \cbox{\includegraphics[width=0.05\linewidth]{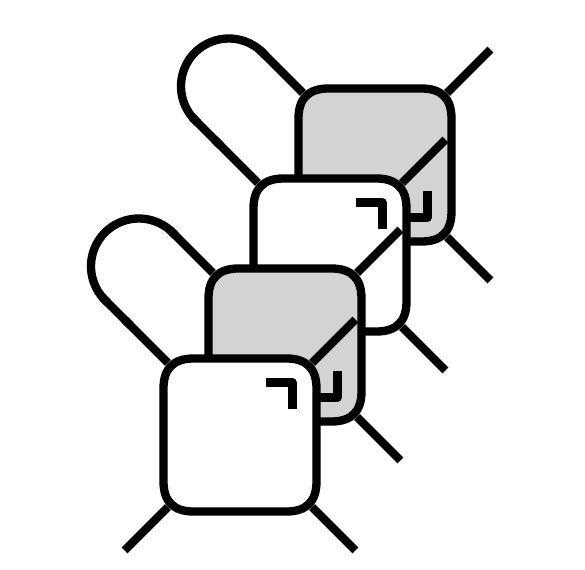}} := \cbox{\includegraphics[width=0.04\linewidth]{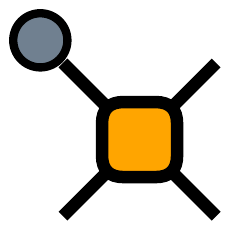}}\,,\quad
	\cbox{\includegraphics[width=0.05\linewidth]{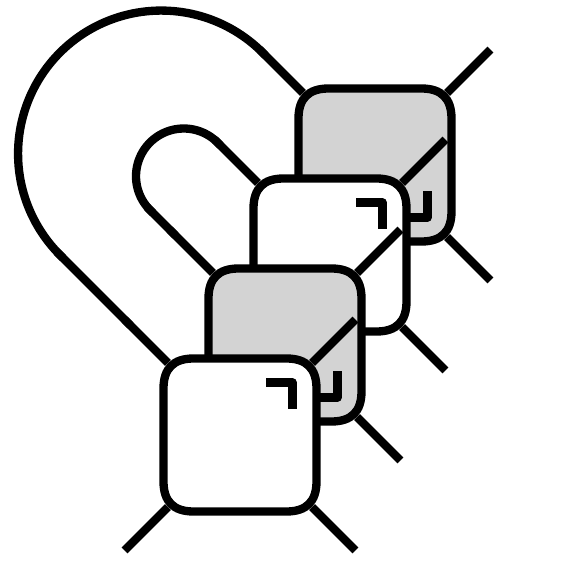}} := \cbox{\includegraphics[width=0.04\linewidth]{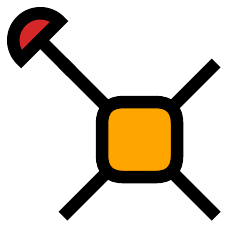}}\,.
	\label{eq:gatedef}
}
The unitarity of the $w$-gates, in terms of these diagrams can be represented as 
\eq{
	\cbox{\includegraphics[width=0.04\linewidth]{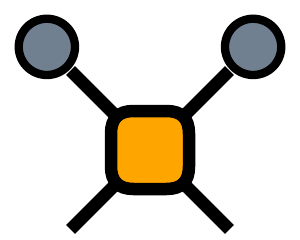}} \!=\! \cbox{\includegraphics[width=0.04\linewidth]{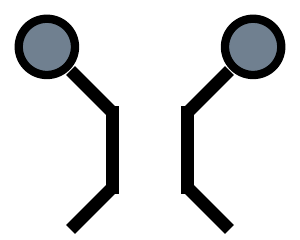}},
	\cbox{\includegraphics[width=0.04\linewidth,angle=180]{4layer-cont1-top.pdf}} \!=\! \cbox{\includegraphics[width=0.04\linewidth,angle=180]{id1-top.pdf}},
	\cbox{\includegraphics[width=0.04\linewidth]{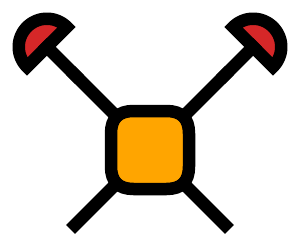}} \!=\! \cbox{\includegraphics[width=0.04\linewidth]{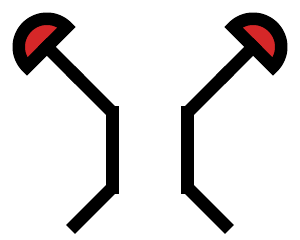}},
	\cbox{\includegraphics[width=0.04\linewidth,angle=180]{4layer-cont2-top.pdf}} \!=\! \cbox{\includegraphics[width=0.04\linewidth,angle=180]{id2-top.pdf}}\,,
	\label{eq:unitary}
}
whereas dual unitarity leads to 
\eq{
	\cbox{\includegraphics[width=0.04\linewidth,angle=90]{4layer-cont1-top.pdf}} = \cbox{\includegraphics[width=0.04\linewidth,angle=90]{id1-top.pdf}}\,,
    \cbox{\includegraphics[width=0.04\linewidth,angle=270]{4layer-cont1-top.pdf}} = \cbox{\includegraphics[width=0.04\linewidth,angle=270]{id1-top.pdf}}\,,
	\cbox{\includegraphics[width=0.04\linewidth,angle=90]{4layer-cont2-top.pdf}} = \cbox{\includegraphics[width=0.04\linewidth,angle=90]{id2-top.pdf}}\,,
	\cbox{\includegraphics[width=0.04\linewidth,angle=270]{4layer-cont2-top.pdf}} = \cbox{\includegraphics[width=0.04\linewidth,angle=270]{id2-top.pdf}}\,.
	\label{eq:dualunitary}
}
Finally note that, since each leg in the replicated picture contains four indices, there are multiple ways of contracting two such legs which yield the following scalars,
\eq{
	\cbox{\includegraphics[width=0.04\linewidth]{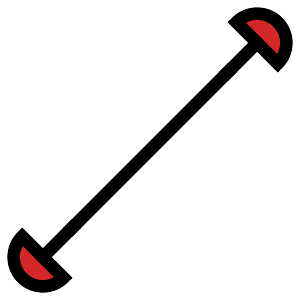}} = 4 = \cbox{\includegraphics[width=0.04\linewidth]{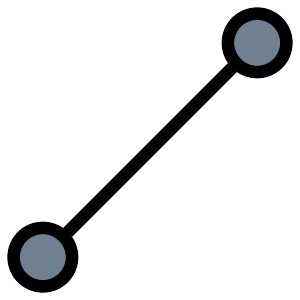}}\,,\quad \cbox{\includegraphics[width=0.04\linewidth]{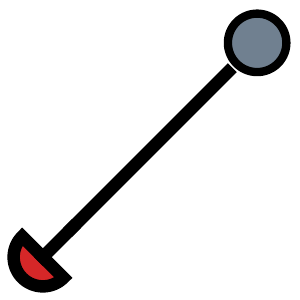}} = 2\,.
    \label{eq:id-rule}
}
With this diagrammatic notation, the eigenstate correlation in Eq.~\ref{eq:eigcorr-gen-t} for an arbitrary choice of subsystems, $X$ and $Y$, can be expressed as
\eq{
	F^{XY}(t) = \cbox{\includegraphics[width=0.2\linewidth]{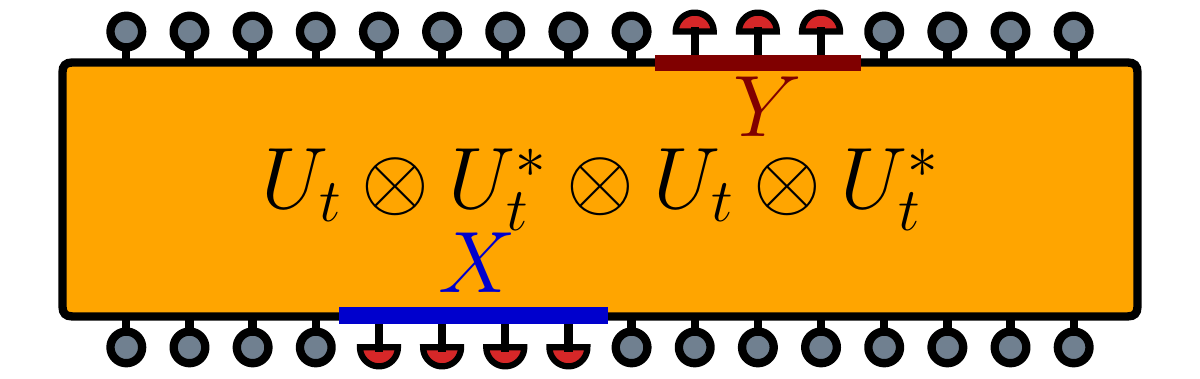}}\,.
    \label{eq:FXY-gen}
}
In the following, we will consider $U_t$ to be made up of a brickwork circuit and use the rules in Eq.~\ref{eq:unitary}--\ref{eq:id-rule} to derive the results presented in the main text.

Consider the case, shown in Fig.~\ref{fig:lightcones}(a) where $X$ and $Y$ are single sites separated by a distance $2t+2d$. In this case $F^{XY}(t)$ can be expressed diagrammatically as 
\eq{
F^{XY}(t) &= 
\cbox{\includegraphics[width=0.2\linewidth]{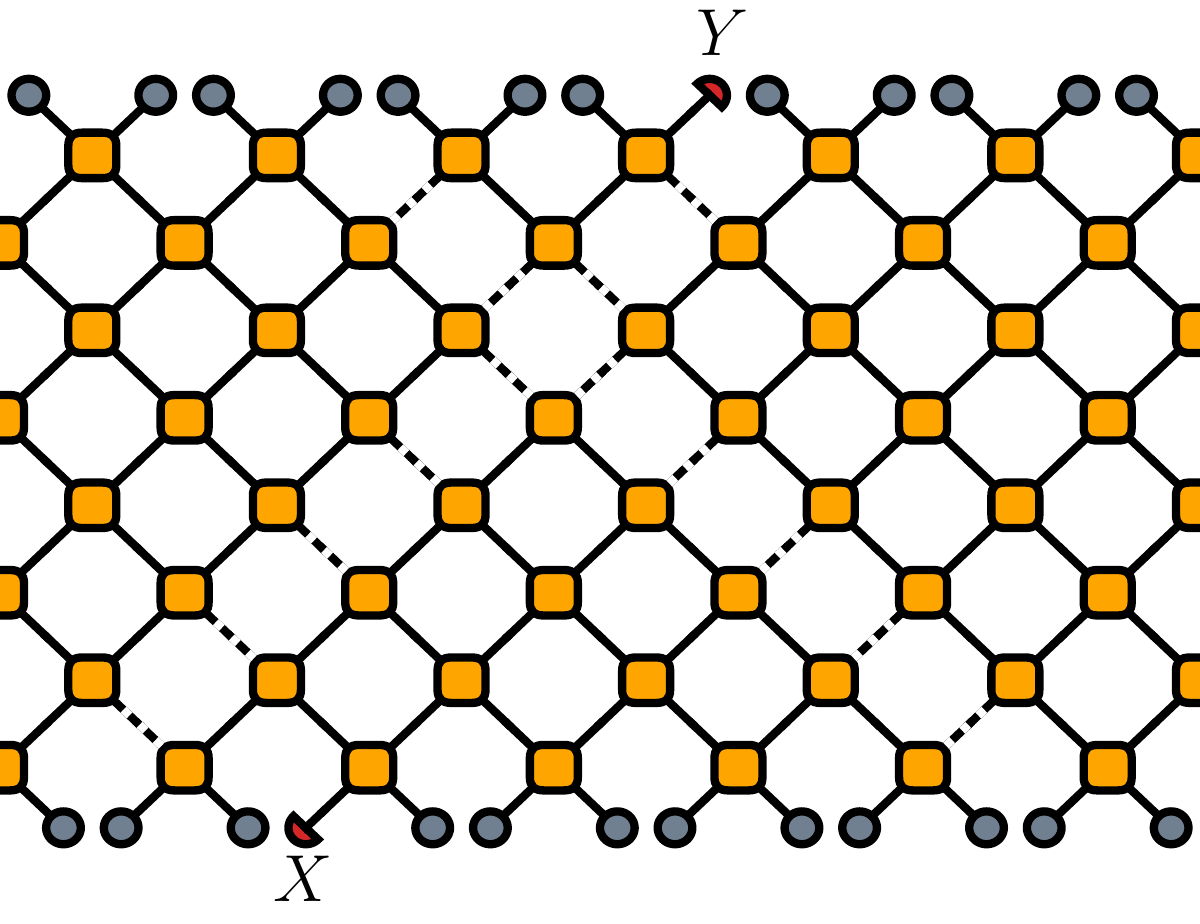}}\nonumber\\&=
\begin{cases}
\cbox{\includegraphics[width=0.2\linewidth]{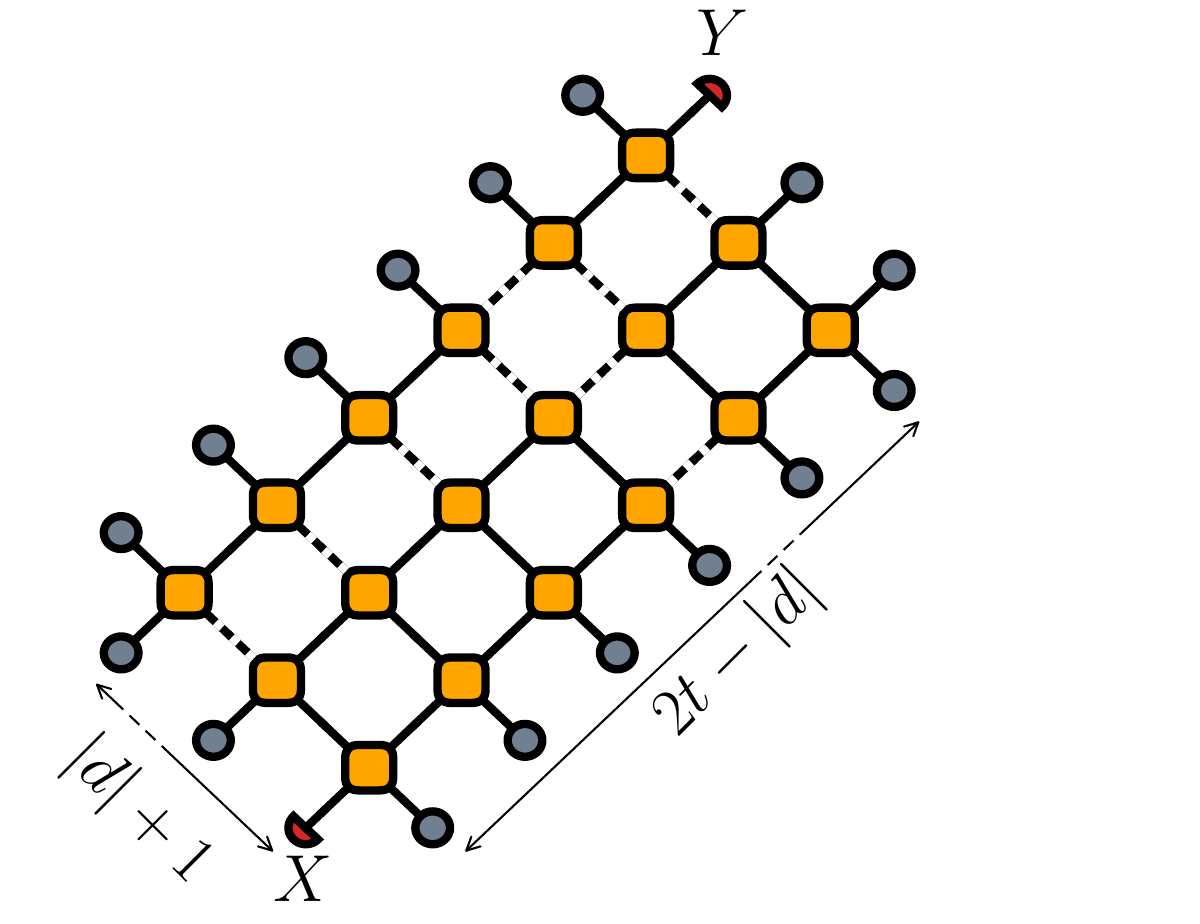}}\!\!\!\!\!\!\!\!\!\!\!\!\!\!\!\times
\left(
\cbox{\includegraphics[width=0.03\linewidth]{cont4c.pdf}}
\right)^{L-2t-1}\!\!\!\,;&d\leq 0\\
\left(
\cbox{\includegraphics[width=0.03\linewidth]{cont4c.pdf}}
\right)^{L-2}\times 
\left(
\cbox{\includegraphics[width=0.03\linewidth]{cont2.pdf}}
\right)^{2}\,;&d>0 
\end{cases}\nonumber\\
&=
\begin{cases}
\cbox{\includegraphics[width=0.18\linewidth]{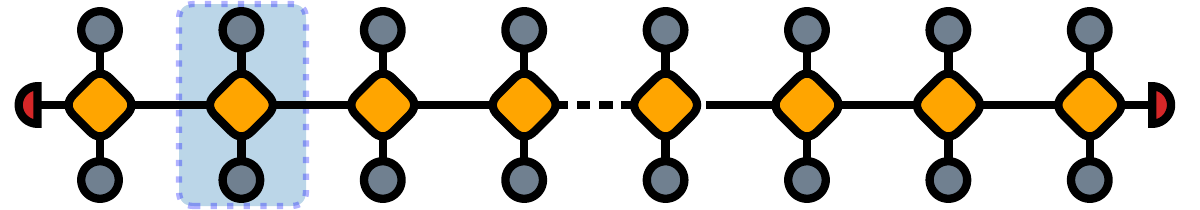}}\times \left(
\cbox{\includegraphics[width=0.03\linewidth]{cont4c.pdf}}
\right)^{L-2t-1}\,; &d=0\\
\left(
\cbox{\includegraphics[width=0.03\linewidth]{cont4c.pdf}}
\right)^{L-2}\times 
\left(
\cbox{\includegraphics[width=0.03\linewidth]{cont2.pdf}}
\right)^{2}\,;&d\neq 0
\end{cases}\,,
\label{eq:FXYt-local-circs}
}
where we used the rules from unitarity, \eqref{eq:unitary}, in going from the first line to the second, and those from dual unitarity, \eqref{eq:dualunitary}, in deriving the third line.
The diagram for $d=0$ in the equation above makes the one-dimensional tensor network structure explicit.
The key point however is that the tensors in the sequence ($w_{i_X,i_X+1},w_{i_X+1,i_X+2},\cdots$) are all different and hence independent. As such, they can be averaged over independently such that the one-dimensional network has the structure of the averaged transfer matrix (indicated by the shaded box) being applied repeatedly.
The averaged transfer matrix has eigenvalues $4$ and $4\Lambda = 4[2-\cos(4J)]/3$ (see Sec.~\ref{suppsec:F_XY local} for details) as mentioned in the main text, which leads directly to the result in Eq.~\ref{eq:FXYt-local-res} for the average eigenstate correlation $\braket{F^{XY}(t)}$.

Turning to $F^{X{\overline{Y}}}(t)$, again for $X$ and $Y$ being single sites, the circuit diagram for the eigenstate correlation is
\eq{
F^{X\overline{Y}}(t) &= 
\cbox{\includegraphics[width=0.2\linewidth]{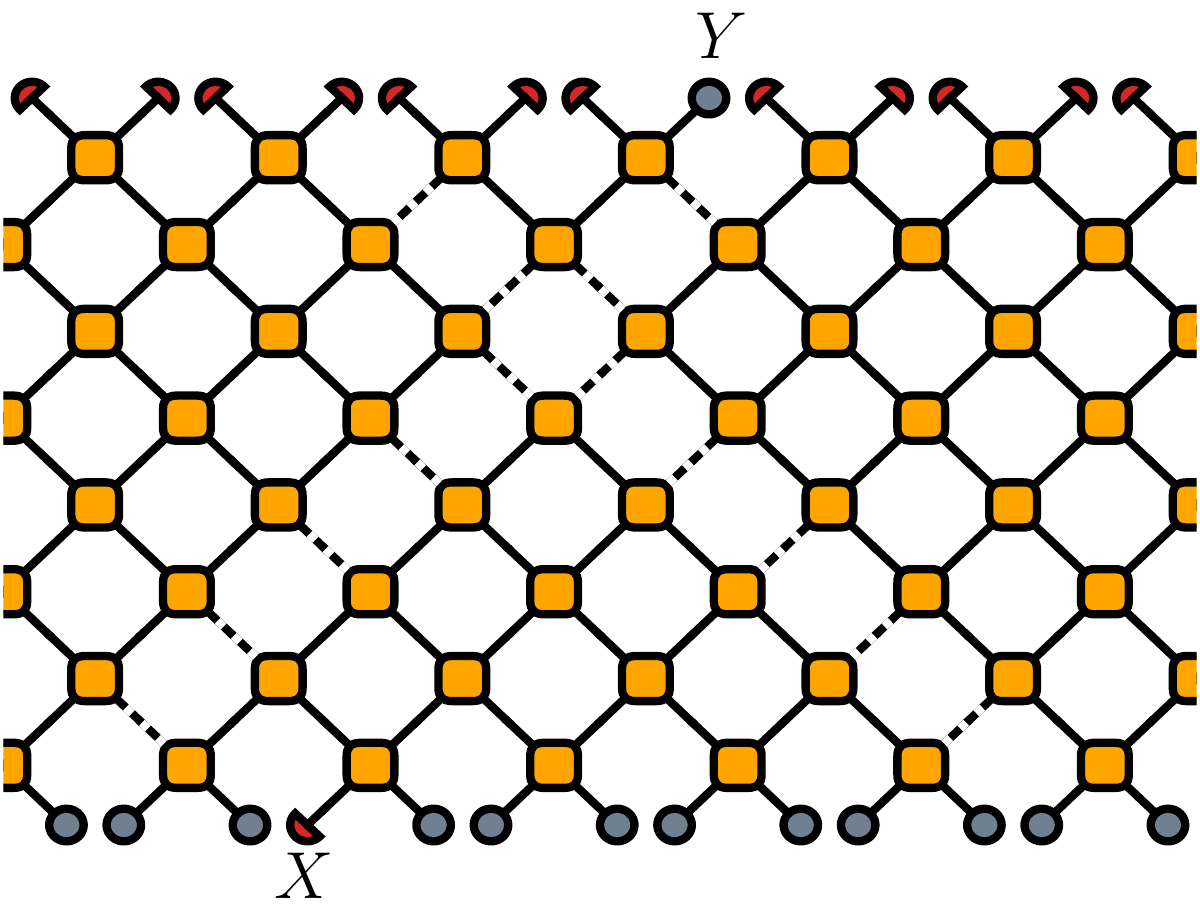}}\nonumber\\&=
\begin{cases}
\cbox{\includegraphics[width=0.2\linewidth]{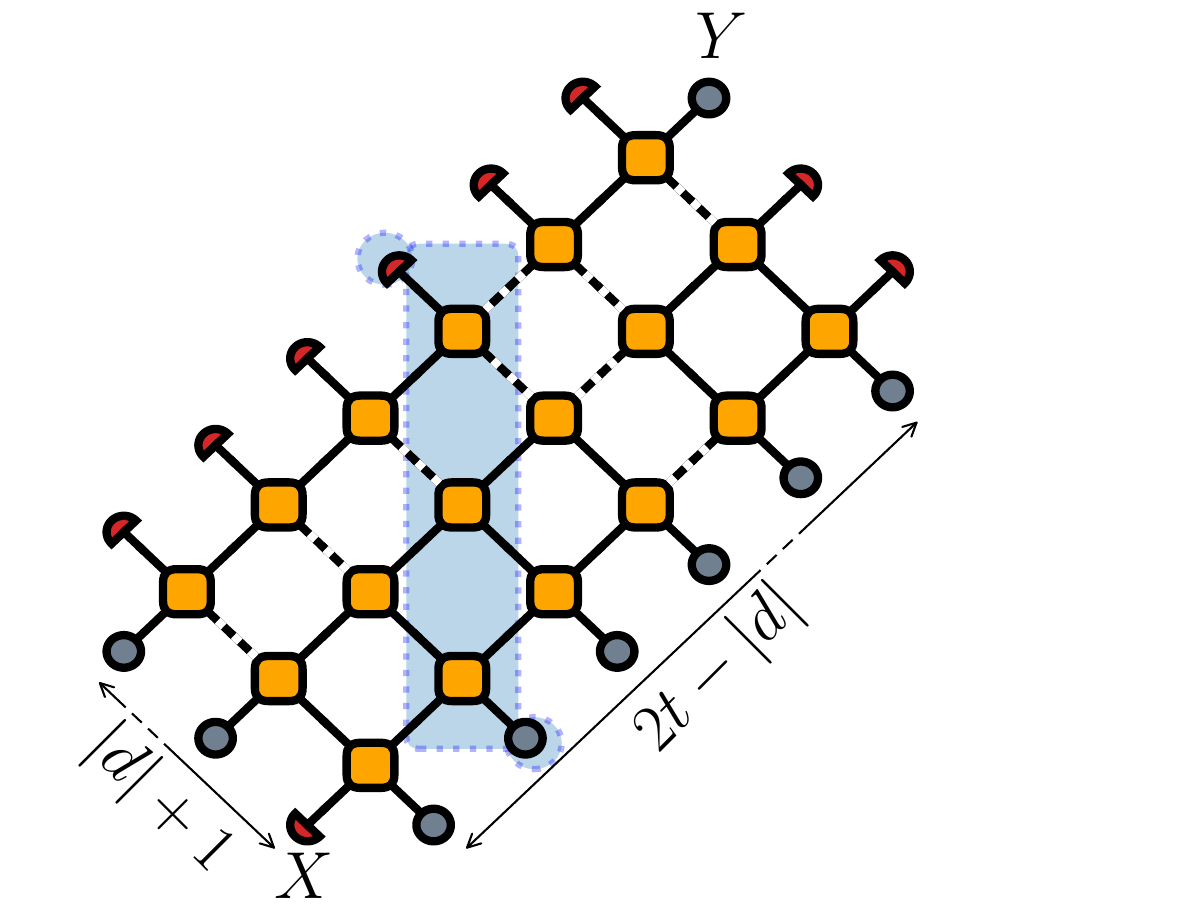}}\!\!\!\!\!\!\!\!\!\times \left(\cbox{\includegraphics[width=0.03\linewidth]{cont2.pdf}}\right)^{L-2t-1}\,;&d \leq 0\\
\left(\cbox{\includegraphics[width=0.03\linewidth]{cont2.pdf}}\right)^{L-2}\times \left(\cbox{\includegraphics[width=0.03\linewidth]{cont4.pdf}}\right)\times \left(\cbox{\includegraphics[width=0.03\linewidth]{cont4c.pdf}}\right)\,; &d>0
\end{cases}\,,
\label{eq:FXYbart-local-circs}
}
where we only use unitarity as dual-unitarity does not lead to any extra simplification except for the case of $d=0$ where the rules in Eq.~\ref{eq:dualunitary} lead to $F^{X\overline{Y}}(t)=2^L$. 
A key point to notice in the circuit diagram above is that, since we are working with a Floquet system, all the gates acting between a given pair of sites are the same, and hence we cannot average over each of the yellow gates individually.
\new{In the case of the commonly considered random DU circuits~\cite{bertini2018exactSFF,bertini2019entanglementspreading,bertini2019exactcorrelations,claeys2020maximumvelocity,bertini2025exactlysolvablemanybodydynamics}, each of the (folded) gates can be averaged over individually which leads to rather simple transfer matrix for any dynamical correlation. Importantly, the dimensionality of the transfer matrix is fixed, which significantly simplifies the computation of the extremal eigenvalue of the transfer matrix which controls the temporal behaviour of the correlations at late times.}
Instead, for the Floquet DU circuits considered here, we need to average over all the gates acting between a given pair of sites simultaneously.
This constitutes the averaged transfer matrix, shown by the blue shaded box again, which acts effectively on 4 copies of $2|d|+1$ spins. Hence, written as a rank-2 tensor, the averaged transfer matrix has a dimension of $2^{4(2|d|+1)}$. 
\new{This renders the computation of the extremal eigenvalue of the transfer matrix which controls the temporal behaviour of the correlations significantly more challenging than the case of random DU circuits.}
We analyse this transfer matrix numerically, details of which are presented in Sec.~\ref{suppsec:FXYbarlocal}.
The analysis shows that the leading eigenvalue of the averaged transfer matrix is 2 whereas the subleading eigenvalue, $\Gamma_d$  depends on $d$, which eventually yields the result in Eq.~\ref{eq:FXYbart-local-res}.


Next we consider the case where $X$ and $Y$ are macroscopic subsystems separated by a distance $r=2t+2d$ (see Fig.~\ref{fig:lightcones}(b)). 
In this case, $F^{XY}(t)$ has the diagrammatic representation
\eq{
F^{XY}(t) &= 
\cbox{\includegraphics[width=0.2\linewidth]{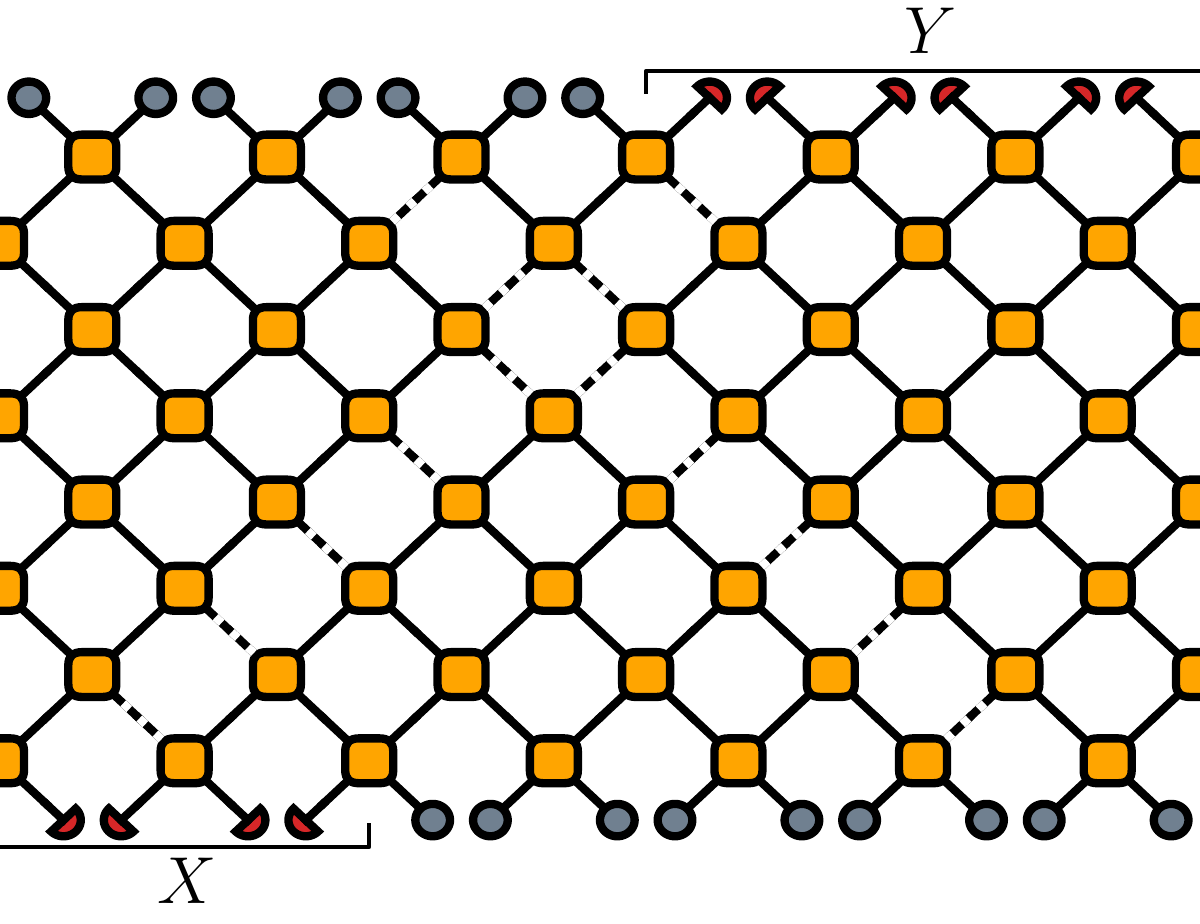}}\nonumber\\
&=
\begin{cases}
\cbox{\includegraphics[width=0.2\linewidth]{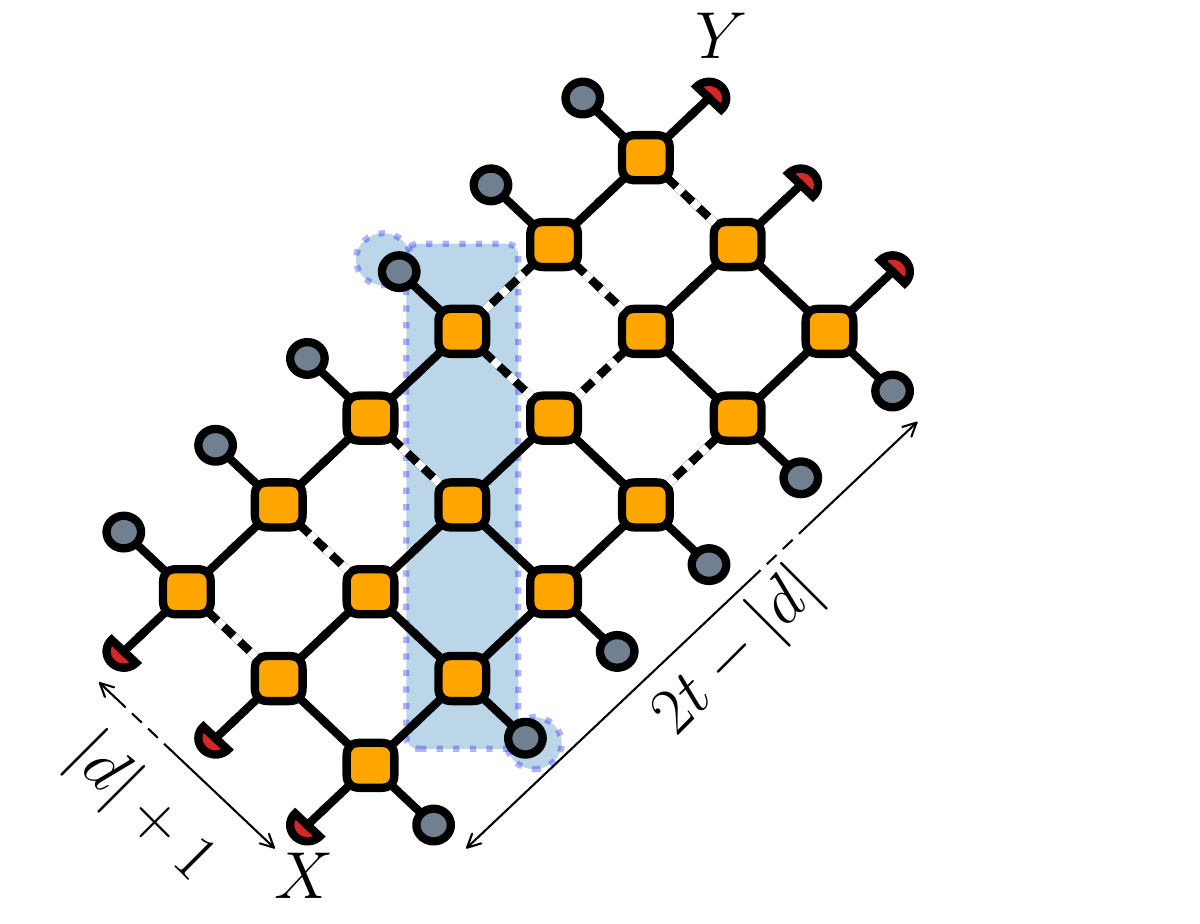}}\!\!\!\!\!\!\!\!\!\times \left(\cbox{\includegraphics[width=0.03\linewidth]{cont2.pdf}}\right)^{L-2t-1}\,;&d\leq 0\\
\left(\cbox{\includegraphics[width=0.03\linewidth]{cont2.pdf}}\right)^{|X|+|Y|}
\left(\cbox{\includegraphics[width=0.03\linewidth]{cont4c.pdf}}\right)^{L-|X|-|Y|}\,;&d>0
\end{cases}\,,
\label{eq:FXYt-nonlocal-circs}
}
where again dual-unitary does not lead to any extra simplification, and the effective transfer matrix's anatomy (shown again with the blued shaded box) is very similar to that for $F^{X\overline{Y}}(t)$ with $X$ and $Y$ single sites (as in Eq.~\ref{eq:FXYbart-local-circs}) except for the different contractions at the ends; this again necessitates numerical analyses of the transfer matrix.
Note however, that for $d=0$, the effective transfer matrix is the same one as for $F^{XY}$ with $X$ and $Y$ single sites as in Eq.~\ref{eq:FXYt-local-circs}. It therefore has non-zero eigenvalues of $4$ and $4\Lambda$. As evinced by the results presented in Sec.~\ref{suppsec:FXYtnonlocal}, these two eigenvalues continue to the leading and the first subleading ones for larger values of $d$ as well. 
This leads to the result in Eq.~\ref{eq:FXYt-nonlocal-res} where the rate of exponential decay in time of $F^{X\overline{Y}}(t)$ is $2|\ln \Lambda|$.
Additional results in the Sec.~\ref{suppsec:FXYtnonlocal} also unambiguously fix the prefactors of the $\Lambda^{2t}$ decay in Eq.~\ref{eq:FXYt-nonlocal-res}.

Turning to $F^{X\overline{Y}}(t)$ for this setting, the circuit diagram looks like
\eq{
&F^{X\overline{Y}}(t) = 
\cbox{\includegraphics[width=0.2\linewidth]{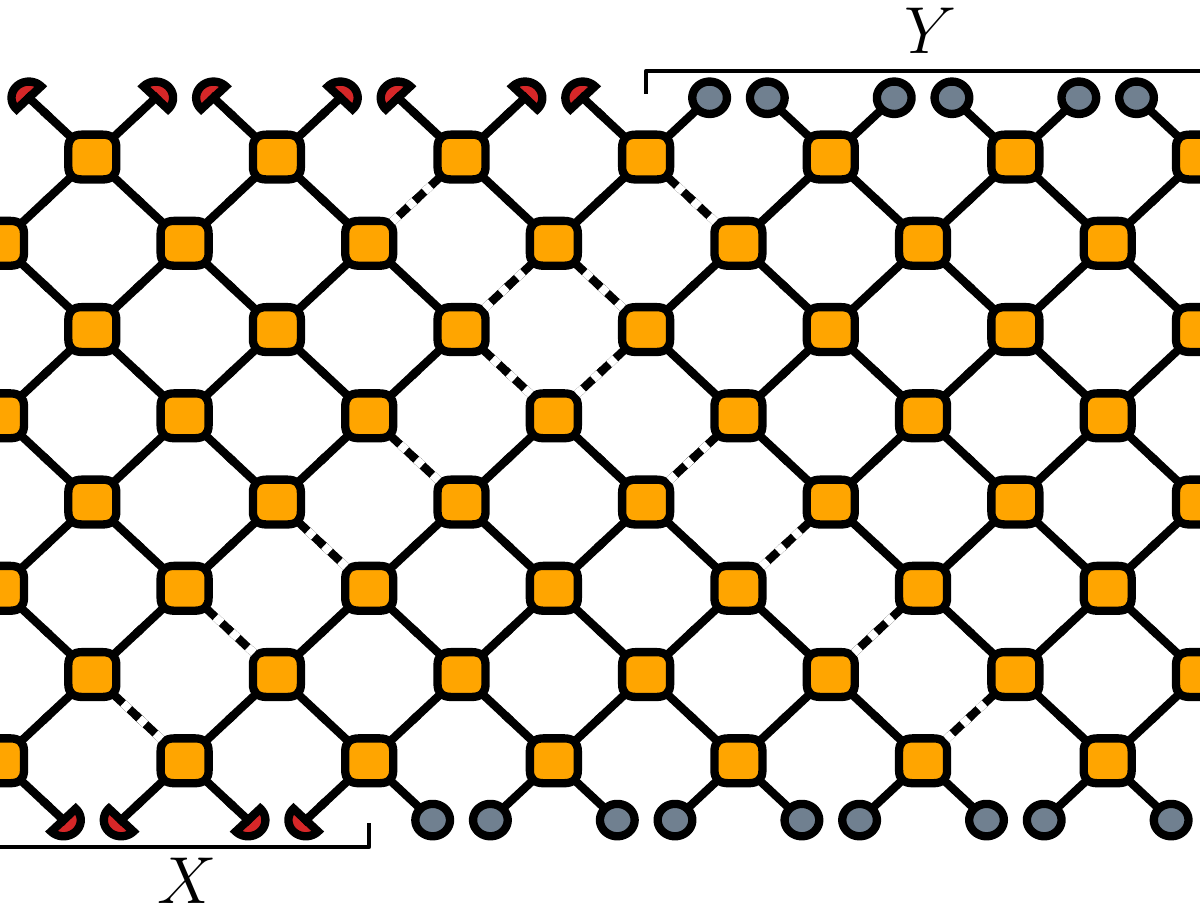}}\nonumber\\
&=
\begin{cases}
\cbox{\includegraphics[width=0.2\linewidth]{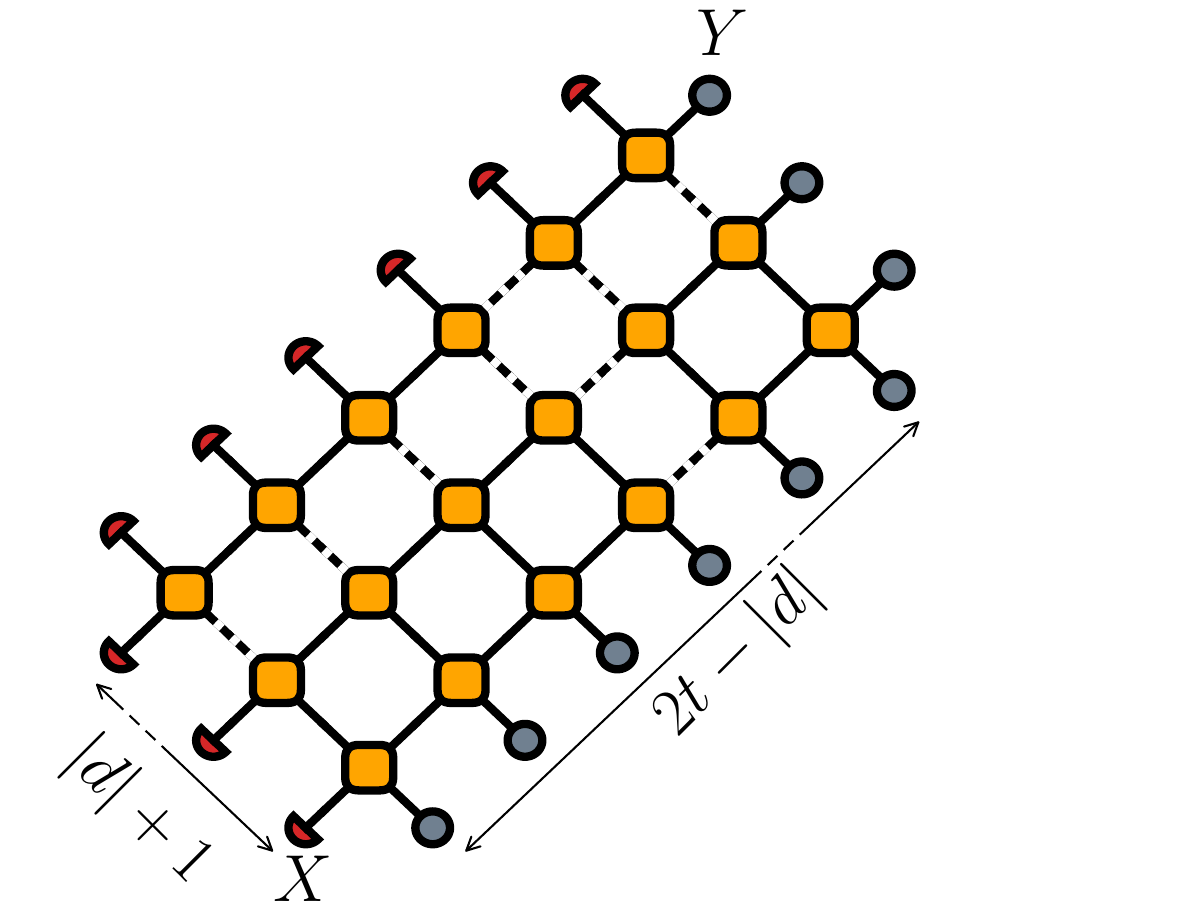}}\!\!\!\!\!\!\!\!\!\!\!\!  
\frac{\left(\cbox{\includegraphics[width=0.03\linewidth]{cont4.pdf}}\right)^{|X|}\left(\cbox{\includegraphics[width=0.03\linewidth]{cont4c.pdf}}\right)^{|Y|}}{\left(\cbox{\includegraphics[width=0.03\linewidth]{cont4.pdf}}\cbox{\includegraphics[width=0.03\linewidth]{cont4c.pdf}}\right)^{|d|+1}}
\,;&d\leq 0\\
\left(\cbox{\includegraphics[width=0.03\linewidth]{cont4.pdf}}\right)^{|X|}
\left(\cbox{\includegraphics[width=0.03\linewidth]{cont4c.pdf}}\right)^{|Y|}
\left(\cbox{\includegraphics[width=0.03\linewidth]{cont2.pdf}}\right)^{L-|X|-|Y|}\,;&d>0
\end{cases}\nonumber\\
&=
\begin{cases}
\left(\cbox{\includegraphics[width=0.03\linewidth]{cont2.pdf}}\right)^{2t+1}\left(\cbox{\includegraphics[width=0.03\linewidth]{cont4.pdf}}\right)^{|X|-|d|-1}
\left(\cbox{\includegraphics[width=0.03\linewidth]{cont4c.pdf}}\right)^{|Y|-|d|-1}
\,;&d\leq 0\\
\left(\cbox{\includegraphics[width=0.03\linewidth]{cont4.pdf}}\right)^{|X|}
\left(\cbox{\includegraphics[width=0.03\linewidth]{cont4c.pdf}}\right)^{|Y|}
\left(\cbox{\includegraphics[width=0.03\linewidth]{cont2.pdf}}\right)^{L-|X|-|Y|}\,;&d>0
 \end{cases}\,,\nonumber
}
where in going from the first line to the second, we used unitarity whereas we used dual-unitarity in going to the third line.
In this case, the circuit-diagrammatic rules are sufficient to completely reduce the eigenstate correlations 
\eq{
F^{X\overline{Y}}(t) = \begin{cases}
        2^{2t+1}\times 4^{|X|+|Y|-2|d|-2}\,;& d\leq 0\\
        2^{L+|X|+|Y|}\,;& d>0
\end{cases}\,.
}
Using the above result in the relation between the eigenstate correlation and the opMI, $I_2^{X\overline{Y}}$, in Eq.~\ref{eq:opMI-FXY}, and the definition of $\Delta_{I_2^{X\overline{Y}(U_t)}}$ defined in the main text, we straightforwardly obtain the result in Eq.~\ref{eq:FXYbart-nonlocal-res}.

\section{$F^{XY}(t)$ for local $X$ and $Y$ on the lightcone} \label{suppsec:F_XY local}

In this section, we discuss the details of the computation of $F^{XY}(t)$ for the case when $X$ and $Y$ are single sites which lie on the lightcone, $i_Y=i_X+2t$.
The key point to note here is that for this case, the $F^{XY}(t)$ boils down to appropriate contractions of an one-dimensional tensor network as
\eq{
F^{XY}(t) = \cbox{\includegraphics[width=0.18\linewidth]{FXY-tmat.pdf}}\times \left(
\cbox{\includegraphics[width=0.03\linewidth]{cont4c.pdf}}
\right)^{L-2t-1}\,.
\label{eq: FXY-local-transfer-mat}}
More importantly, the sequence of tensors in the one-dimensional network are built successively out of different tensors, $w_{X,X+1},w_{X+1,X+2},\cdots$ and so on, such that they are independent as the single-site Haar random unitaries, $u_\pm,v_\pm$ parametrising the $w$-gates, as in Eq.~\ref{eq:w-gate-def}, are independently chosen for each of them.
This implies that one can independently average over the single-site Haar random gates to obtain a translation-invariant averaged gate
\eq{
\mathbb{E}\left(\!\!\!\!\!\!\!\!\!\!\!\!\!\!\!\!\!\!\cbox{\includegraphics[width=0.18\linewidth,angle=-45]{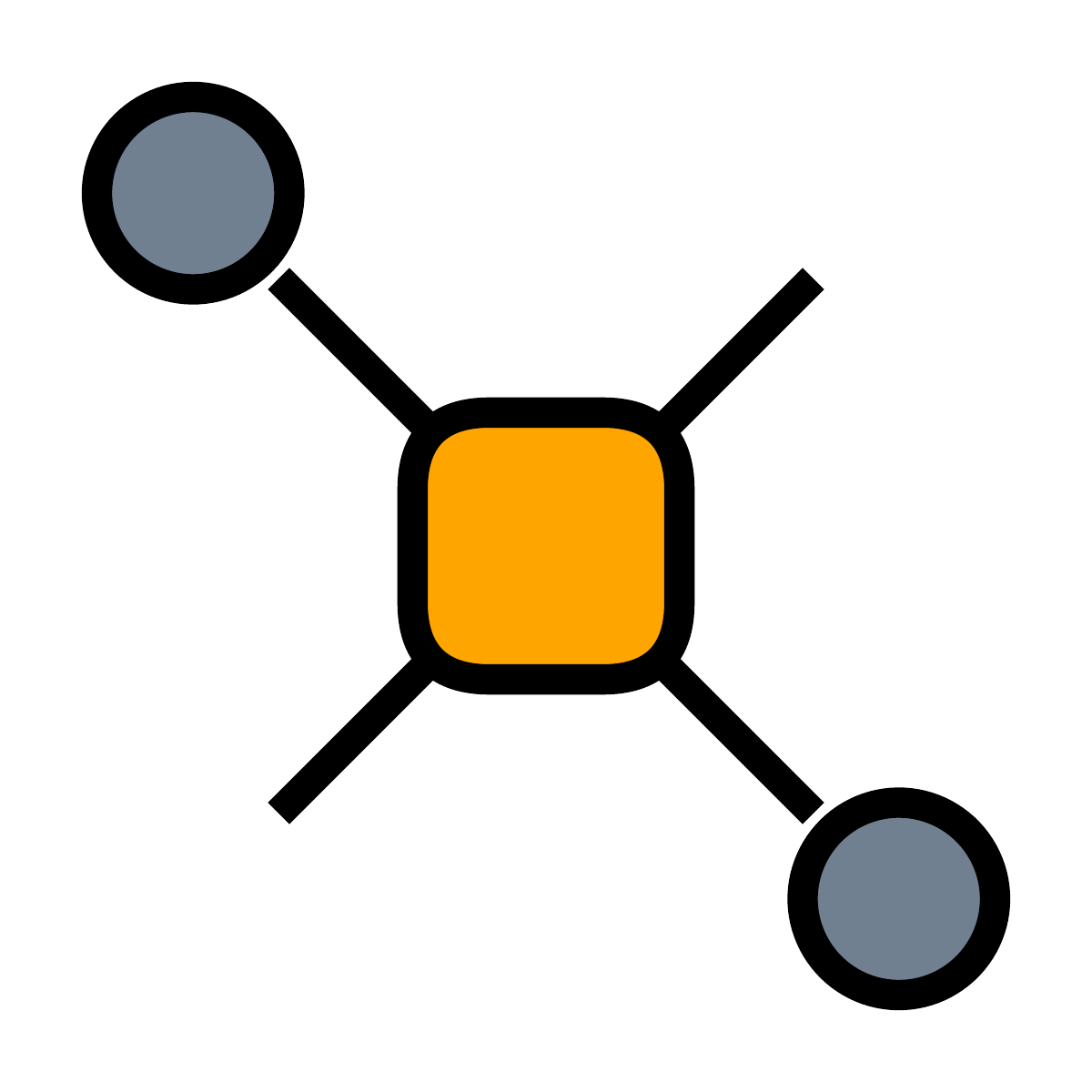}}\!\!\!\!\!\!\!\!\!\!\!\!\!\!\!\!\!\!\right)_{u_\pm,v_\pm}=
\mathbb{E}\left(\!\!\!\!\!\!\!\!\!\!\!\!\cbox{\includegraphics[width=0.2\linewidth,angle=-45]{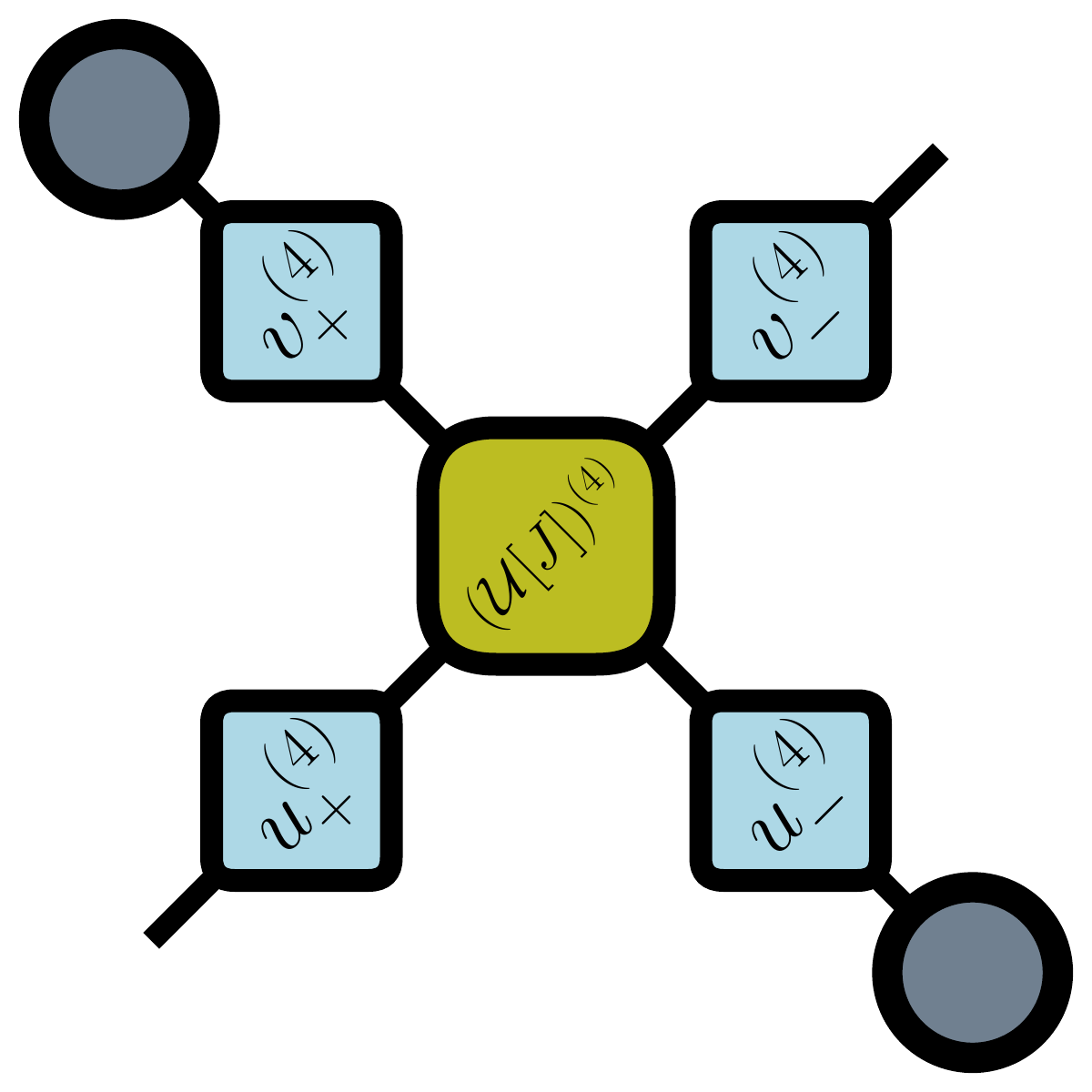}}\!\!\!\!\!\!\!\!\!\!\!\!\right)_{u_\pm,v_\pm}=
\!\!\!\!\!\!\!\!\!\!\!\!\cbox{\includegraphics[width=0.15\linewidth,angle=-45]{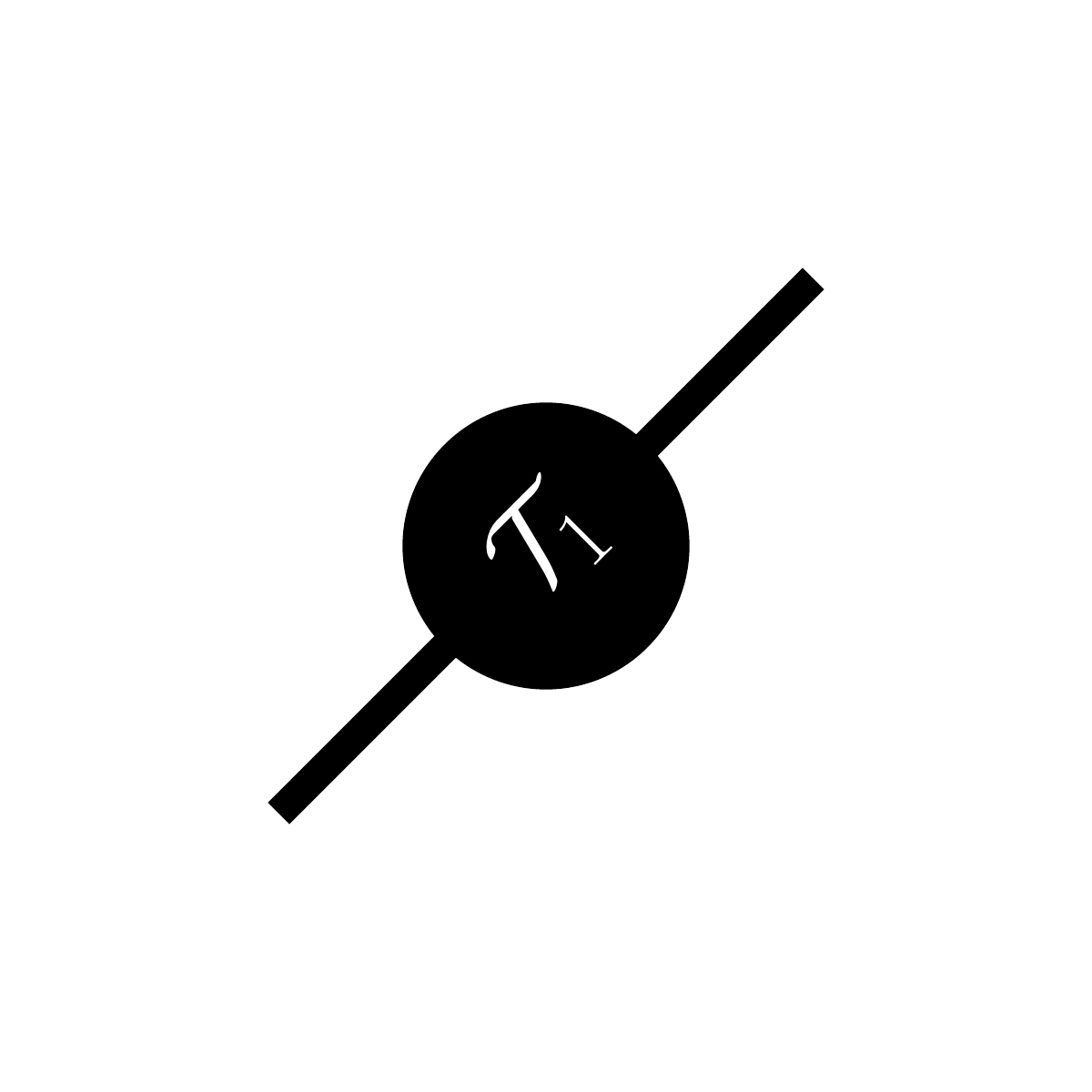}}\!\!\!\!\!\!\,,
\label{eq:T1-def}
}
where the notation $({\cal U}[J])^{(4)}\equiv {\cal U}[J]\otimes ({\cal U}[J])^\ast \otimes {\cal U}[J]\otimes ({\cal U}[J])^\ast $ and the operator ${\cal T}_1$, denoted by the black circle acting on the four-copy Hilbert space denotes the averaged operator.

With this notation, $F^{XY}(t)$ for $d=0$ can be expressed as 
\eq{
F^{XY}(t) = 
\cbox{\includegraphics[width=0.06\linewidth]{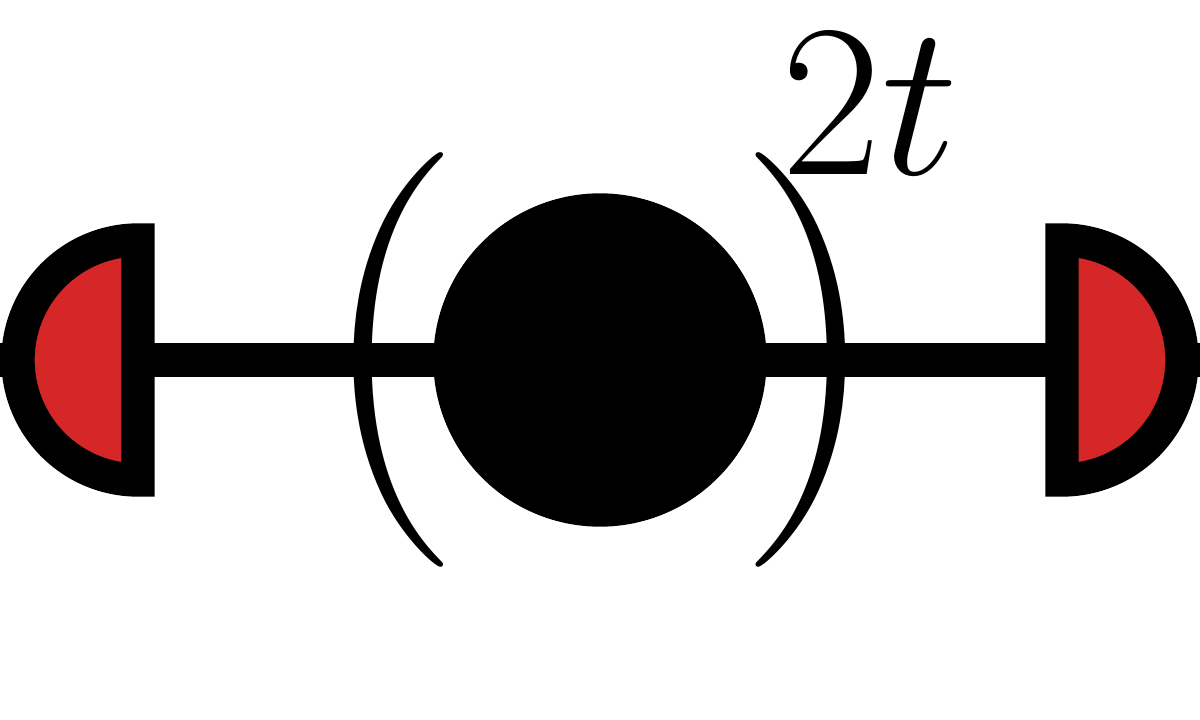}}\times
\left(
\cbox{\includegraphics[width=0.03\linewidth]{cont4c.pdf}}
\right)^{L-2t-1}\,.
\label{eq:FXYlocal-tmat-sm}
}
Note that each leg in the ${\cal T}_1$ operator carries 4 indices with each index taking $q=2$ values. Denoting the the incoming indices as $(s_1,s_2,s_3,s_4)$ and the outgoing ones as $(s_1^\prime,s_2^\prime,s_3^\prime,s_4^\prime)$, the tensor ${\cal T}_1$ can be written as
\eq{
{\cal T}_1 = \sum_{\substack{s_1,s_2,s_3,s_4 \\ s_1^\prime,s_2^\prime,s_3^\prime,s_4^\prime}}\ket{s_1^\prime,s_2^\prime,s_3^\prime,s_4^\prime}\bra{s_1,s_2,s_3,s_4}({\cal T}_1)_{s_1,s_2,s_3,s_4}^{s_1^\prime,s_2^\prime,s_3^\prime,s_4^\prime}\,,
}
where the elements of the tensor are given by 
\eq{
({\cal T}_1)_{s_1,s_2,s_3,s_4}^{s_1^\prime,s_2^\prime,s_3^\prime,s_4^\prime} = & \delta_{s_1s_2}\delta_{s_3s_4}\delta_{s_1^\prime s_2^\prime}\delta_{s_3^\prime s_4^\prime}\times \frac{1}{9}(10 + 2\sin^2 2J) +\nonumber\\
& \delta_{s_1s_2}\delta_{s_3s_4}\delta_{s_1^\prime s_4^\prime}\delta_{s_2^\prime s_3^\prime}\times \frac{-2}{9}(1 + 2\sin^2 2J) +\nonumber\\
&\delta_{s_1s_4}\delta_{s_2s_3}\delta_{s_1^\prime s_2^\prime}\delta_{s_3^\prime s_4^\prime}\times \frac{-2}{9}(1 + 2\sin^2 2J)\nonumber+\\
&\delta_{s_1s_4}\delta_{s_2s_3}\delta_{s_1^\prime s_4^\prime}\delta_{s_2^\prime s_3^\prime}\times \frac{4}{9}(1 + 2\sin^2 2J)\,.
}
Combining the indices $(s_1,s_2,s_3,s_4)$ into a composite index $S$ and similarly  $(s_1^\prime,s_2^\prime,s_3^\prime,s_4^\prime)$ into $S^\prime$, the tensor ${\cal T}_1$ can be expressed as a $16\times 16 $ matrix which can be readily diagonalised.
Diagonalisation leads to two non-zero eigenvalues, $e_0=4$ and $e_1 = 4(2-\cos 4J)/3$ and the respective eigenvectors denoted by $\ket{e_{0/1}}$. The tensor ${\cal T}_1^{2t}$ can therefore be written as 
\eq{
{\cal T}_1^{2t} = 4^{2t}\left[\ket{e_0}\bra{e_0}+ \left(\frac{2-\cos 4J}{3}\right)^{2t}\ket{e_1}\bra{e_1}\right]\,.
\label{eq:T1-eigs}}
Given any tensor $(\mathfrak{T})_{s_1,s_2,s_3,s_4}^{s_1^\prime,s_2^\prime,s_3^\prime,s_4^\prime}$, the red semicircular contractions on boths ends (such as in Eq.~\ref{eq:FXYlocal-tmat-sm}) simply yields the scalar $\sum_{s_1,s_2,s_1^\prime,s_2^\prime}(\mathfrak{T})_{s_1,s_2,s_2,s_1}^{s_1^\prime,s_2^\prime,s_2^\prime,s_1^\prime}$. Using this relation for the projectors onto the eigenvectors $\ket{e_{0/1}}\bra{e_{0/1}}$, we have from Eq.~\ref{eq:FXYlocal-tmat-sm},
\eq{
\braket{F^{XY}(t)} =  4^{2t}\left[1+ 3\left(\frac{2-\cos 4J}{3}\right)^{2t}\right]\times 4^{L-2t-1} =4^{L-1}\left[1+ 3\left(\frac{2-\cos 4J}{3}\right)^{2t}\right]\,,
}
which is exactly the result in Eq.~\ref{eq:FXYt-local-res}.

\section{$\braket{F^{X\overline{Y}}(t)}$ for local $X$ and $Y$ \label{suppsec:FXYbarlocal} }

In this section, we present the details of computing $\braket{F^{X\overline{Y}}(t)}$, where $X$ and $Y$ are local and are at a distance $i_Y-i_X = 2t + 2d$ as illustrated in Fig.~\ref{fig:lightcones}(a). Specifically, we consider the case of $d<0$ which corresponds to $Y$ lying inside the lightcone, in which case the $\braket{F^{X\overline{Y}}(t)}$ is given by
\eq{
F^{X\overline{Y}}(t) = \cbox{\includegraphics[width=0.3\linewidth]{FXYbar2.pdf}}\!\!\!\!\!\!\!\!\!\times \left(\cbox{\includegraphics[width=0.05\linewidth]{cont2.pdf}}\right)^{L-2t-1}
\label{eq:FXYbar-local-circ}
}
The yellow gates acting on a given pair of sites are identical across the circuit.
This means that the averaged transfer matrix has to be constructed out of averaging the vertical stack of $|d|+1$ identical gates as
\eq{
\mathbb{E}\left(\cbox{\includegraphics[width=0.05\linewidth]{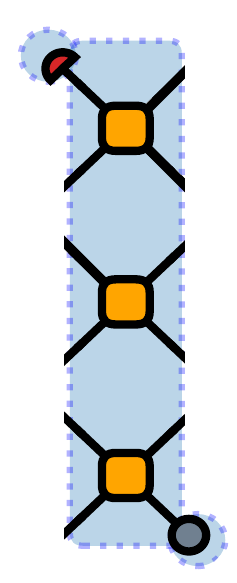}}\right) = \cbox{\includegraphics[width=0.07\linewidth]{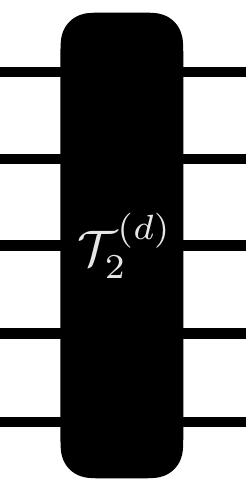}}\,,
}
where we denote this averaged transfer matrix as ${\cal T}_2^{(d)}$.
Note that the above diagram makes it clear that the averaged transfer matrix, written as rank-2 tensor has dimensions of $2^{4(2|d|+1)}$ as each of the legs acts on four copies of the circuit.
From Eq.~\ref{eq:FXYbar-local-circ}, it follows that ${\cal T}_2^{(d)}$ needs to be applied $2t-2|d|$ times to obtain $\braket{F^{X\overline{Y}}(t)}$ as
\eq{
\braket{F^{X\overline{Y}}(t)} = \cbox{\includegraphics[width=0.2\linewidth]{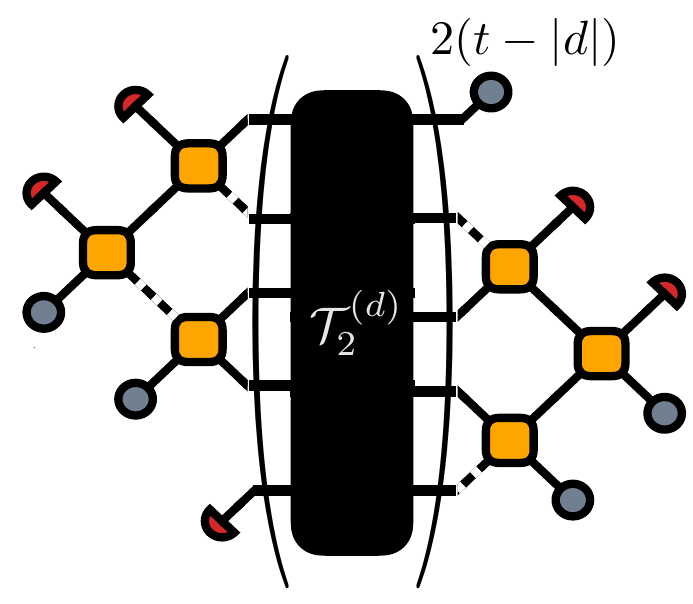}}\times \left(\cbox{\includegraphics[width=0.05\linewidth]{cont2.pdf}}\right)^{L-2t-1}\,.
\label{eq:FXYbar-tmat}
}
We find that the transfer matrix ${\cal T}_2^{(d)}$ is Hermitian which allows us to decompose it as
\eq{
{\cal T}_2^{(d)} = \sum_{\lambda_d}\lambda_d\ket{\lambda_d}\bra{\lambda_d}\,,
}
which in turn means that Eq.~\ref{eq:FXYbar-tmat} can be expressed as 
\eq{
\braket{F^{X\overline{Y}}(t)} = 2^{L-2t-1}\times \sum_{\lambda_d}\left[\lambda_d^{2(t-|d|)}\times\underbrace{\cbox{\includegraphics[width=0.25\linewidth]{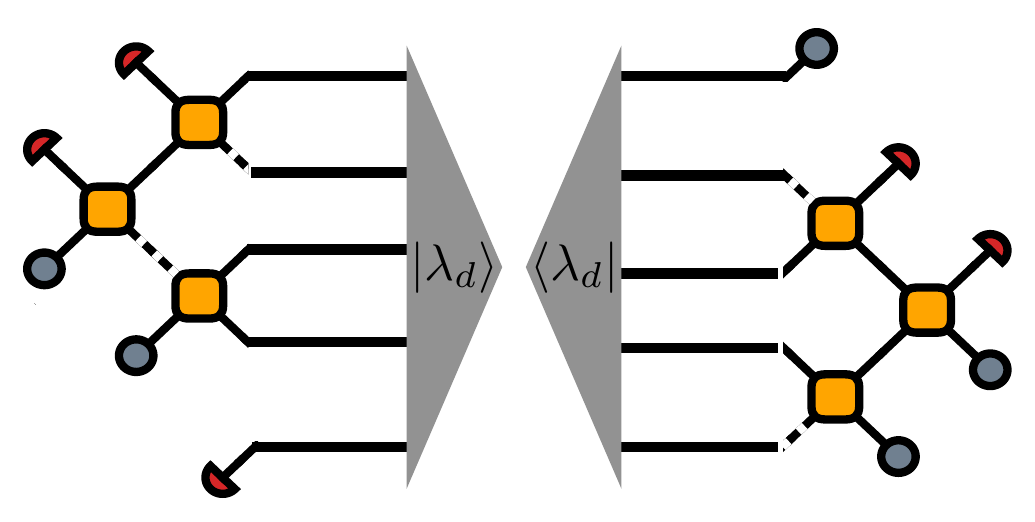}}}_{C_{\lambda_d}}\right]\,.
\label{eq:FXYbar-local-eigen}
}
A numerical analyses of the averaged transfer matrix ${\cal T}_2^{(d)}$ shows that its largest eigenvalue is $\lambda_d^{\rm max} = 2$ and both $\lambda_d^{\rm max}$ as well as the subleading eigenvalues are degenerate.
The form in Eq.~\ref{eq:FXYbar-local-eigen} suggests that at long times the average eigenstate correlation can be written as 
\eq{
\braket{F^{X\overline{Y}}(t)} = 2^{L-2t-1}\times\left[(\lambda_d^{\rm max})^{2(t-|d|)}\nu_d^{\rm max} + \nu_d (2\Gamma_d)^{2(t-|d|)}+\cdots\right]\,.
}
where $\nu_d^{\rm max}$ is the total contribution of the $C_{\lambda_d}$s coming from all the degenerate eigenvectors corresponding to eigenvalue $\lambda_d^{\rm max}$, $2\Gamma_d$ is the largest subleading eigenvalue with a non-vanishing eigenvector contribution $\nu_d$ where $\nu_d$ is the total eigenvector contribution from all the degenerate eigenvectors corresponding to eigenvalue $2\Gamma_d$.

\begin{figure}
\centering
\includegraphics[width=0.7\linewidth]{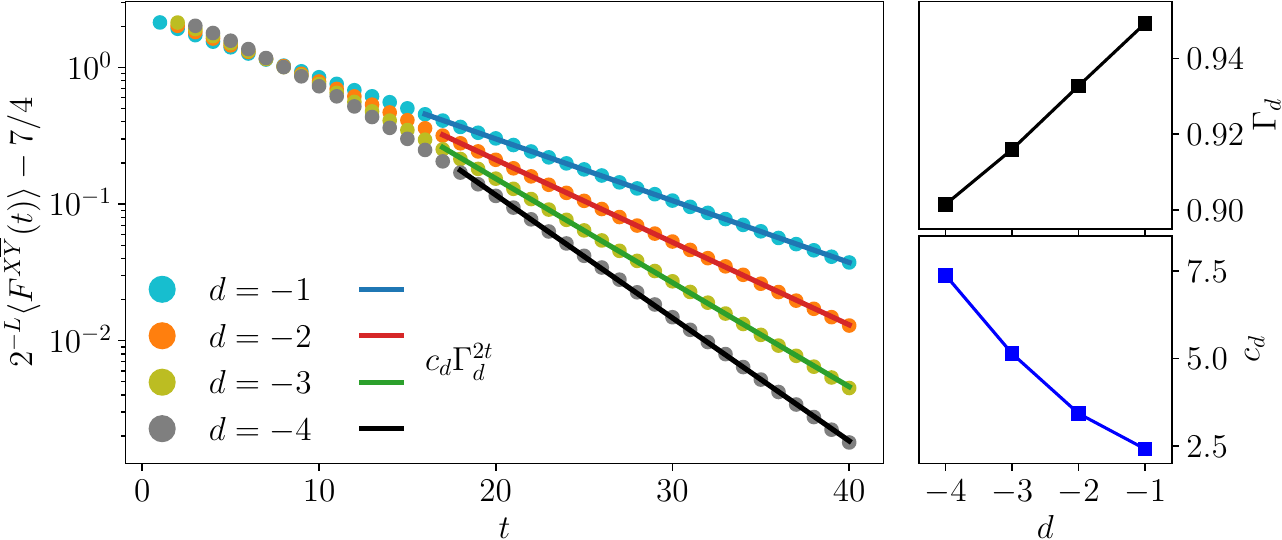}
\caption{Left: Numerical results for the time dependence of $F_{X\overline{Y}}(t)$ for $J = 0.5$ and for various values of $d$, showing exponential decay. Solid lines show the fit $c_d \Gamma_d^{2t}$. Right: Variation of $\Gamma_d$ and $c_d$ with $d$.}
\label{fig:F_XYbar-local-decay}
\end{figure}

In Fig.~\ref{fig:F_XY_non-local-decay} we show the behaviour of $\Gamma_d$ with $d$ which shows that the former decreases with increasing $|d|$ which in turn implies that the eigenstate correlation decays faster the further inside the lightcone $Y$.
In addition, we also compute $\nu_d^{\rm max}$ as well as a $\nu_d$ numerically for a few values of $d$. 
We find that $\nu_d^{\rm max} = 7\times 2^{2|d|-1}$ and we define
$c_d = 2\nu_d/7(2\Gamma_d)^{2|d|}$; the latter is also plotted as function of $d$ in Fig.~\ref{fig:F_XYbar-local-decay}.
Putting these together we find the result
$$
\braket{F^{X\overline{Y}}(t)} = 2^L \times 
    \frac{7}{4}\left(1 + c_d \Gamma_d^{2t} + \cdots\right),
$$
which is precisely the result in Eq.~\ref{eq:FXYbart-local-res}.
Numerically evaluating the circuit in Eq.~\ref{eq:FXYbar-local-circ} indeed confirms the above behaviour at late times as shown in Fig.~\ref{fig:F_XYbar-local-decay}.

We next discuss briefly the minor differences in the case where $X$ is on an even site but $Y$ is on an odd site. We again consider $i_Y=i_X + 2t+ 2d$ but now with $d$ the negative of a half-integer.
In this case $F^{X\overline{Y}}(t)$ is given by the circuit diagram 
\eq{
F^{X\overline{Y}}(t) = \cbox{\includegraphics[width=0.3\linewidth]{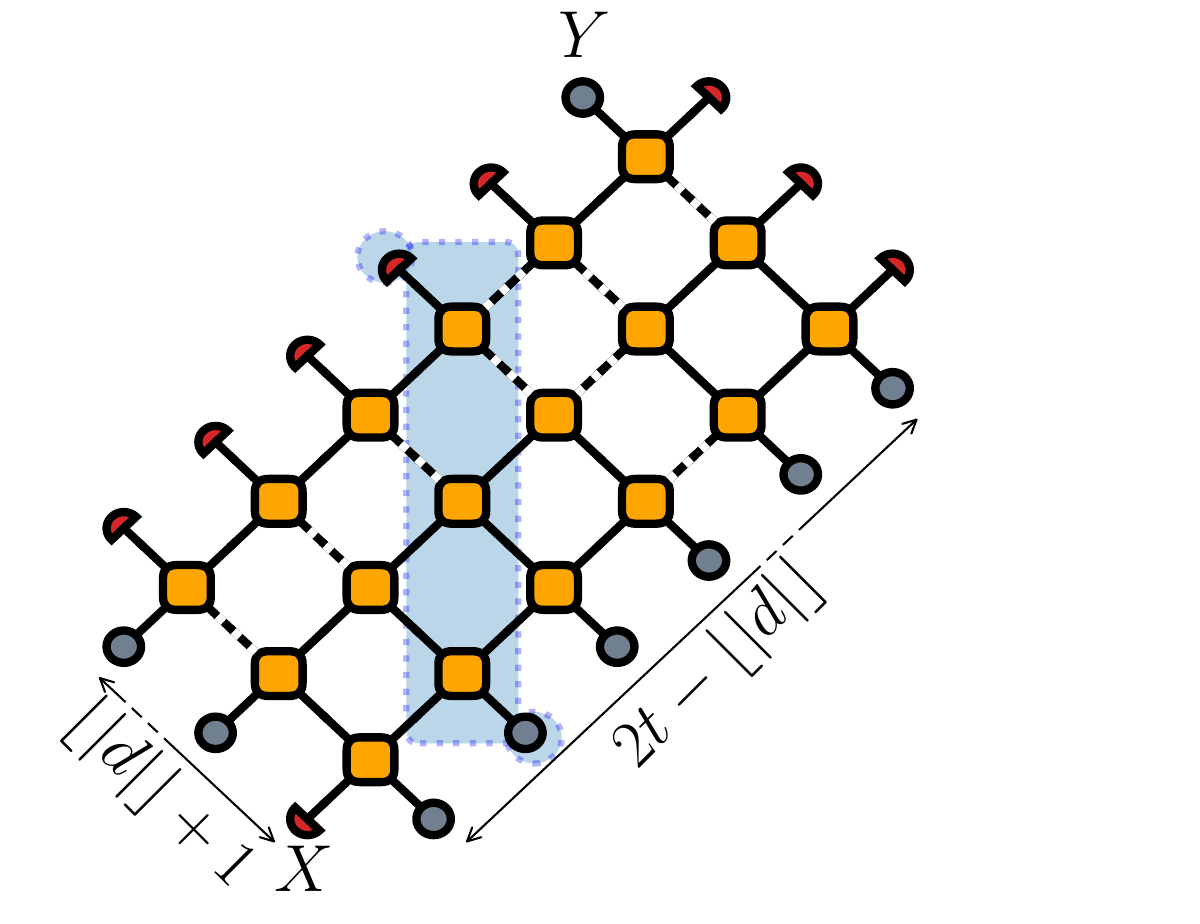}}\!\!\!\!\!\!\!\!\!\times \left(\cbox{\includegraphics[width=0.05\linewidth]{cont2.pdf}}\right)^{L-2t-1}\,,
\label{eq:FXYbar-local-circ-odd-even}
}
which looks rather similar to the case with both $X$ and $Y$ on even sites, Eq.~\ref{eq:FXYbar-local-circ}, but with a very minor difference in the contractions at the top of the circuit near $Y$. To highlight the quantitative difference between the even-even and even-odd case, consider first $d=-1/2$ which has the same circuit geometry as the $d=0$ case modulo the aforementioned difference in the contraction. 
For $d=-1/2$, $F^{X\overline{Y}}(t)$ can be reduced, using the rules from dual-unitarity, to
\eq{
F^{X\overline{Y}}(t)&=
\cbox{\includegraphics[width=0.05\linewidth]{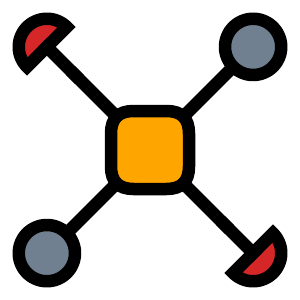}}\times \left(\cbox{\includegraphics[width=0.05\linewidth]{cont2.pdf}}\right)^{L-2}\,.}
Note that the remaining yellow gate is nothing but the same gate appearing in Eq.~\ref{eq: FXY-local-transfer-mat} whose average yields the transfer matrix ${\cal T}_1$ defined in Eq.~\ref{eq:T1-def}. We therefore have,
\eq{
F^{X\overline{Y}}(t)&=2^{L-2}\times 
\cbox{\includegraphics[width=0.05\linewidth]{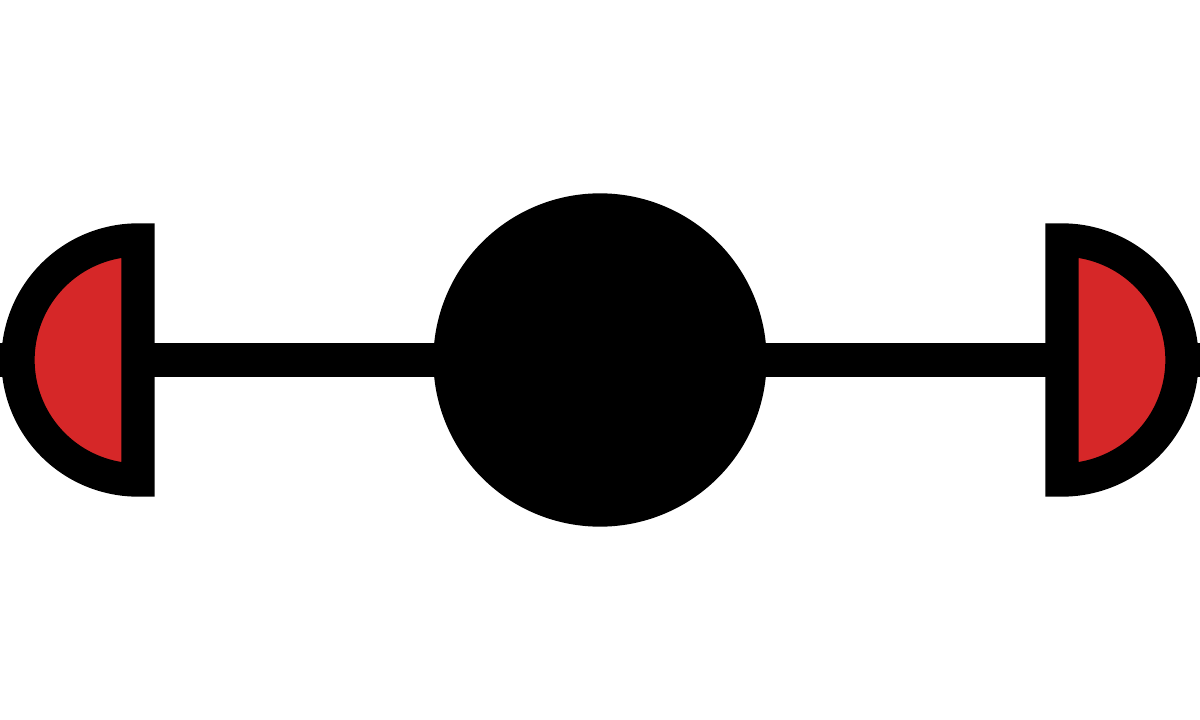}}=2^L \times (3-\cos{4J})\,,
}
where for the last equality we used the properties of the eigenvalues and eigenvectors of ${\cal T}_1$ obtained in \hyperref[suppsec:F_XY local]{Sec.~I}.
Similar to the case of $d=0$, in this case of $d=-1/2$ also, $\braket{F^{X\overline{Y}}(t)}$ is a constant with time. 
However, the value of this constant depends on $J$ explicitly unlike the $d=0$ case where the value is just $2^L$, independent of $J$.

A more important point to note is that the average transfer matrix that emerges for negative half-integer values of $d$ is identical to that of integer values of $d$. 
This is straightforwardly seen from the fact that the blue shaded region (whose average gives us the transfer matrix) is identical in Eq.~\ref{eq:FXYbar-local-circ} and Eq.~\ref{eq:FXYbar-local-circ-odd-even}. 
As such, for negative half-integer values of $d$  we have
\eq{
\braket{F^{X\overline{Y}}(t)} = \cbox{\includegraphics[width=0.35\linewidth]{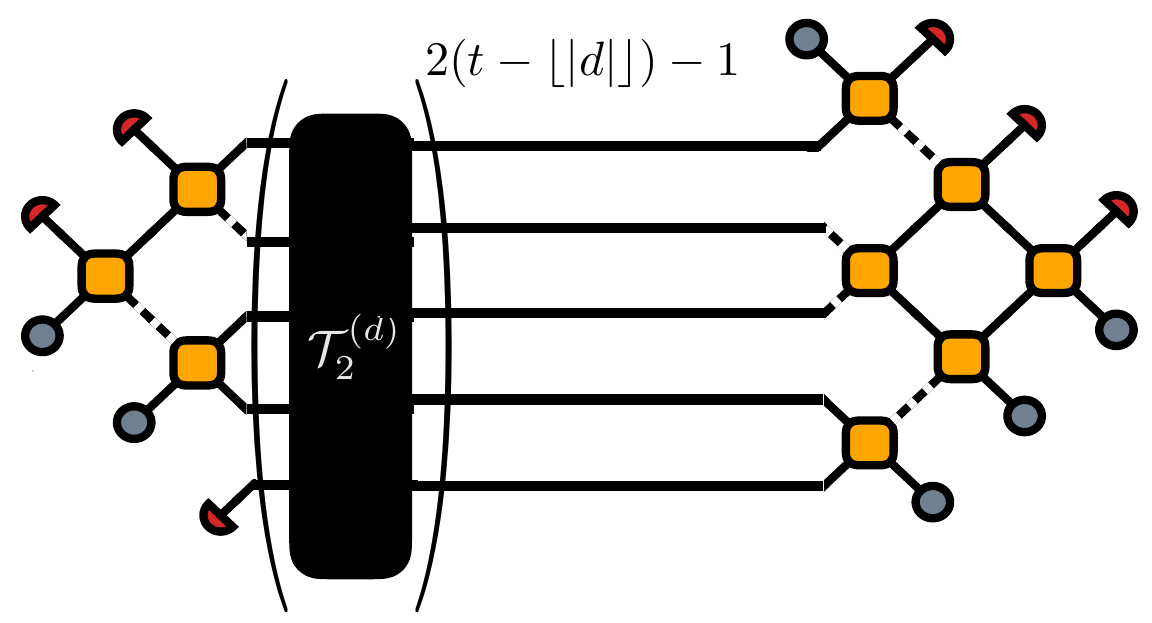}}\times \left(\cbox{\includegraphics[width=0.05\linewidth]{cont2.pdf}}\right)^{L-2t-1}\,,
\label{eq:FXYbar-tmat-oddeven}
}
which is identical to Eq.~\ref{eq:FXYbar-tmat} except for the contractions on the right boundary. 
The latter can only affect the eigenvector contributions.
The exponential decay rate, governed by the first subleading eigenvalue therefore continues to be the same as $2|\ln\Gamma_d|$ whereas the different boundary contractions affect the saturation value at $t\to\infty$ and the prefactor in front of the $\Gamma_d^{2t}$ decay.

\section{$\braket{F^{XY}(t)}$ for macroscopic $X$ and $Y$ \label{suppsec:FXYtnonlocal}}

In this section, we present the details of computing the correlation function $\braket{F^{XY}(t)}$, where $X$ and $Y$ are macroscopic. The setting, shown in Fig.~\ref{fig:lightcones}(b), is such that $Y$ is to the right of $X$ and  the support of $Y$ begins at a distance $r = 2t + 2d$ from the rightmost site of $X$. In this case, $\braket{F^{XY}(t)}$ reduces to,
\eq{
\braket{F^{XY}(t)} = \cbox{\includegraphics[width=0.3\linewidth]{FXY2-nonlocal.pdf}}\!\!\!\!\!\!\!\!\!\times \left(\cbox{\includegraphics[width=0.05\linewidth]{cont2.pdf}}\right)^{L-2t-1}.
\label{eq:FXY-nonlocal-circ}
}

As in the earlier local case, the yellow gates acting on any given pair of sites are identical across the circuit. This allows us to define an averaged transfer matrix, $\mathcal{T}_3^{(d)}$, which acts on four copies of $2|d| + 1$ spins, yielding a Hilbert space of dimension $2^{4(2|d| + 1)}$. This transfer matrix differs from that in the local case due to the symmetric contractions at both closed ends (see the top-left and bottom-right corners in the shaded region of the figure). For $d = 0$, it reduces to the same 1D network discussed in Eq.~\ref{eq: FXY-local-transfer-mat}. The averaged transfer matrix, built from a stack of $|d| + 1$ identical gates, is given by

\eq{
\mathbb{E}\left(\cbox{\includegraphics[width=0.05\linewidth]{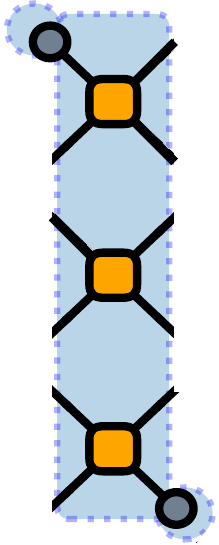}}\right) = \cbox{\includegraphics[width=0.07\linewidth]{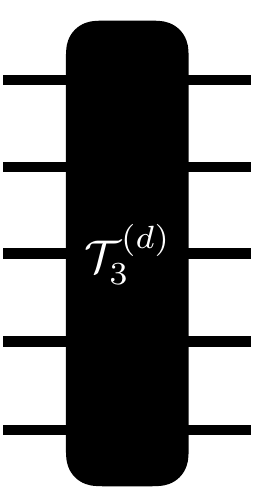}}\,.
}
From Eq.~\ref{eq:FXY-nonlocal-circ}, it follows that the averaged transfer matrix needs to be applied $2t-2|d|$ times to obtain 
\eq{
\braket{F^{XY}(t)} = \cbox{\includegraphics[width=0.2\linewidth]{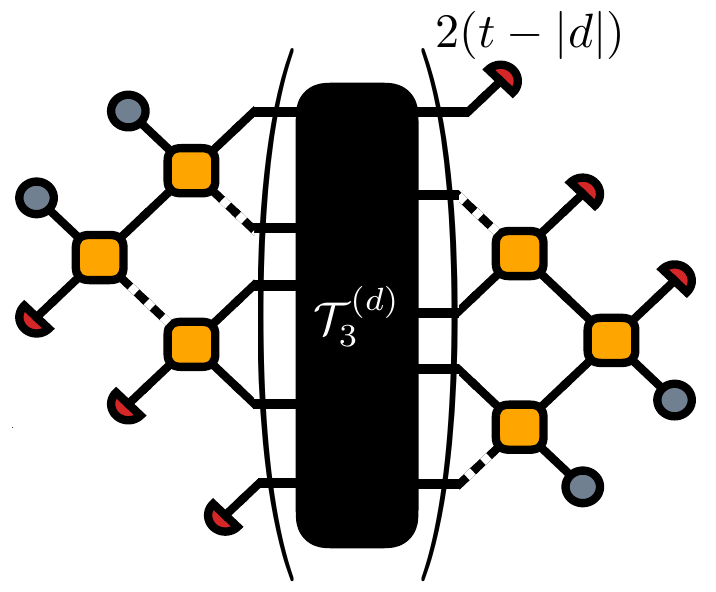}}\times \left(\cbox{\includegraphics[width=0.05\linewidth]{cont2.pdf}}\right)^{L-2t-1}\,.
\label{eq:FXY-nonlocal-tmat}
}
The transfer matrix ${\cal T}_3^{(d)}$ is Hermitian, allowing for a spectral decomposition,
\eq{
{\cal T}_3^{(d)} = \sum_{\lambda_d}\lambda_d\ket{\lambda_d}\bra{\lambda_d}\,,
}
which in turn means that Eq.~\ref{eq:FXY-nonlocal-tmat} can be expressed as 
\eq{
\braket{F^{XY}(t)} = 2^{L-2t-1}\times \sum_{\lambda_d}\left[\lambda_d^{2(t-|d|)}\times\underbrace{\cbox{\includegraphics[width=0.25\linewidth]{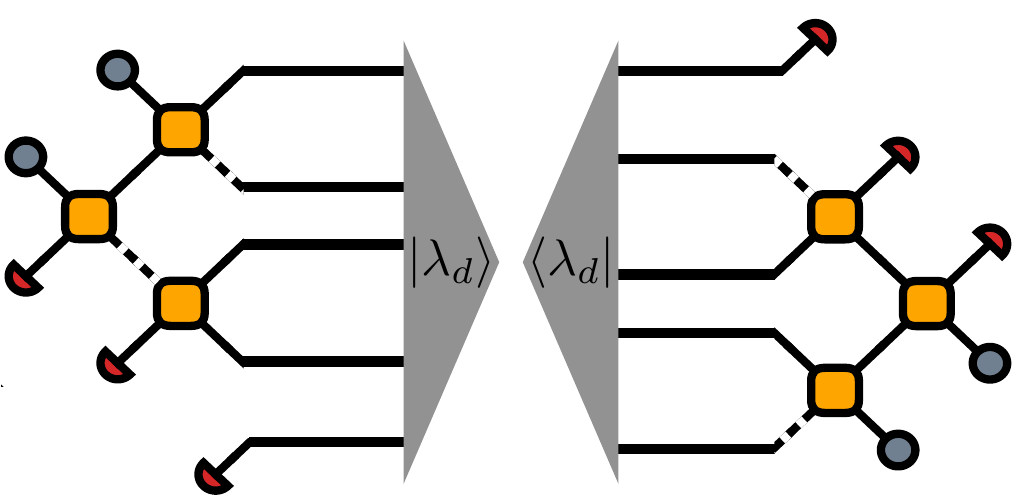}}}_{E_{\lambda_d}}\right]\,.
\label{eq:FXYexten-eigen}
}
with $E_{\lambda_d}$ being the contribution of the eigenstates that have the eigenvalue $\lambda_d$. Here, $E_{\lambda_d}\geqslant 0 , \,\,\forall \,\lambda_d$ necessarily because contractions on the left and right eigenvectors are conjugate to each other. As discussed in the main text, the quantity of interest here is the mutual information,

\eq{\braket{\exp[I_2^{XY}(U_t)]}=\frac{\braket{F^{XY}(t)}}{2^{L+2(t+d)-1}}=2^{-4t+2|d|}\sum_{\lambda_d}\lambda_d^{2t-2|d|}E_{\lambda_d},}

\begin{figure}
    \centering
    \includegraphics[width=0.8\linewidth]{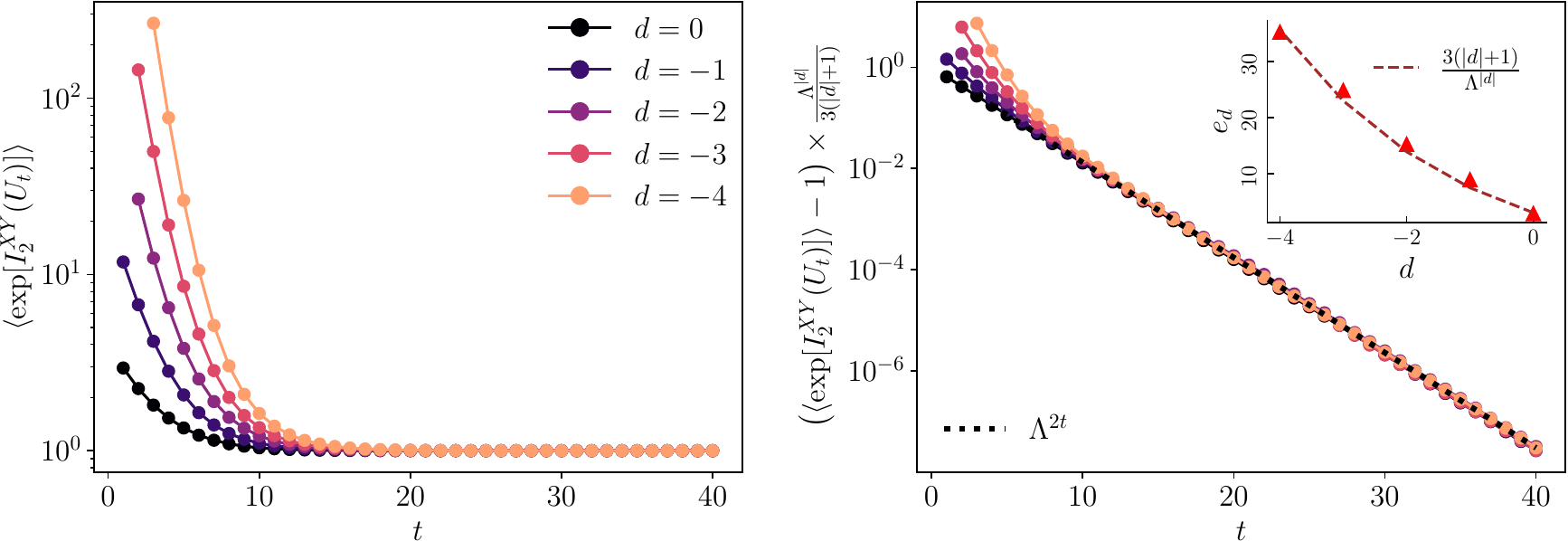}
   \caption{Left: $\braket{\exp[I_2^{XY}(U_t)]}$ as a function of $t$ for various values of $d\leq 0$ for $J=0.5$.
Right: Data collapse of $\braket{\exp[I_2^{XY}(U_t)]}-1$, demonstrating scaling behavior.
The inset shows the plot of $e_d$ versus $d$, showing an exponential growth with $|d|$ with a rate $|\ln\Lambda|$ dressed by a multiplicative correction, linear in $|d|$.}
    \label{fig:F_XY_non-local-decay}
\end{figure}

For $d = 0$, we can relate this to Eq.~\ref{eq:T1-eigs}, where $\mathcal{T}_3^{(0)} = \mathcal{T}_1$ is analytically tractable. Its largest two eigenvalues are $4$ and $4\Lambda$, with $\Lambda = (2 - \cos(4J)) / 3$, giving
\eq{\braket{\exp[I_2^{XY}(U_t)]}=1+ 3\Lambda^{2t}+\cdots\,.}
For $d<0$ the numerical results shown in Fig.~\ref{fig:F_XY_non-local-decay} imply that the saturation value as well as the decay rate is the same as those for $d=0$. 
The latter is a fallout of the fact that the subleading eigenvalue of ${\cal T}_3^{(d)}$ is invariant with $d$. However, the corresponding eigenvector contribution depends on $d$ as $e_d \approx 3(|d| + 1)/\Lambda^{|d|}$ as suggested by the result in the inset to the right panel in Fig.~\ref{fig:F_XY_non-local-decay}, where $e_d$ is again the total contribution from the all the (possibly) degenerate eigenvectors corresponding to eigenvalue $4\Lambda$. 
Putting all of this together, we have
\eq{
\braket{\exp[I_2^{XY}(U_t)]} = 1 + e_d \Lambda^{2t} + \cdots\,;\quad \quad e_d \approx 3(|d| + 1)/\Lambda^{|d|}\,,
}
which is precisely the result in Eq.~\ref{eq:FXYt-nonlocal-res}.

\end{document}